**Article type: Review Article**

# Engineering Molecular Rectification: Mechanisms, Modulation Strategies, and Device Integration

Junnan Guo[1] , Shufan Song[1], Wenhui Fang[1], Jifeng Tang[1], Wenhao Li[1], Weikang Wu[1], Hui Li[1*], Shishen Yan[2*] and Lishu Zhang[1*]

[1]Key Laboratory for Liquid-Solid Structural Evolution and Processing of Materials, Ministry of Education, Shandong University, Jinan 250061, China

[2] School of Physics, Shandong University, Jinan 250100, China

E-mail: lihuilmy@sdu.edu.cn

E-mail: shishenyan@sdu.edu.cn

E-mail: lishu.zhang@sdu.edu.cn

## Abstract

Molecular rectifiers, as prototypical components of molecular electronics, present unique opportunities for pushing device miniaturization to its ultimate limits. Nevertheless, challenges including limited rectification ratios (RR), insufficient robustness, and poor reproducibility impede their practical deployment. To make molecular rectifiers competitive with silicon-based devices, it is important to fully understand the design principles and fabrication methods from both mechanistic and experimental perspectives. By holistically considering the transport mechanisms, modulation strategies, fabrication, characterization techniques, and theoretical simulations, this review provides a comprehensive overview of molecular rectifiers. Representative examples of conceptually significant and high-performance molecular rectifier systems are highlighted to illustrate the relationships between rectification mechanisms, molecular design strategies, and device realization. Building on these discussions, we present an outlook for current bottlenecks and future directions to guide the development of molecular rectifiers. This review aims to serve as both a conceptual framework and a technical reference for researchers working at the intersection of molecular electronics and nanoscale device engineering in the post-CMOS era.

## 1 Introduction

The continuous downscaling of electronic components has pushed conventional silicon-based technology to the edge of its physical and functional limits.[1-4] Once device dimensions fall below the sub-5 nm node, leakage current rises by more than three orders of magnitude compared with the 10 nm process, carrier mobility degrades by over 40% due to short-channel effects, and thermal noise further amplifies signal interference.[5-10] These non-ideal physical phenomena dominate charge transport, undermining both the performance and scalability of traditional complementary metal-oxide-semiconductor (CMOS) architectures. Against this backdrop, molecular electronics has emerged as a promising alternative by employing single molecules or small ensembles as functional units.[11-15] Exploiting intrinsic quantum features such as quantum interference and tunable energy levels, molecular electronics not only enables ultimate miniaturization toward the sub-1 nm regime but also allows the integration of rectification, switching, and memory functionalities within tailored molecular structures.[16-18] This unique capability offers unprecedented opportunities for the miniaturization and diversification of post-Moore electronic systems.

Among various molecular electronic devices, the molecular rectifier stands as both the earliest and the most representative prototype.[19-21] In 1974, Aviram and Ratner proposed the first theoretical model based on a donor-σ-acceptor (D-σ-A) architecture.[12] By exploiting the energy-level offset between donor and acceptor units, they demonstrated unidirectional current flow at the molecular scale. This mechanism departed fundamentally from the doping-dependent rectification of conventional *p-n* junctions. Their pioneering concept not only established the physical foundation of molecular rectifiers but also marked the transition of molecular electronics from visionary concept to scientific discipline. To quantify device performance, the RR was defined as the ratio between the forward current under a given bias and the corresponding reverse current. A higher RR indicates stronger directional selectivity of charge transport and thus greater relevance for practical rectification.

Over the past five decades, both theoretical and experimental studies of molecular rectifiers have advanced significantly, yet achieving device performance comparable to silicon rectifiers has proved elusive. It was not until 2017 that the first rectifier with an RR of $10^5$ was realized through a self-assembled monolayer (SAM)-based junction. [22-24] Subsequent efforts have pushed RR values into the range of $10^6$-$10^8$, but such results are often limited to idealized conditions, falling short of the operational stability and reliability required for practical applications.[25, 26] In most cases, rectification remains modest with RR below $10^3$.[27-29] This limitation can be largely attributed to insufficient asymmetry between molecular orbitals and electrode Fermi levels, as well as the difficulty of simultaneously achieving interfacial tunability and long-term reliability.[30-32] High operating voltages, limited conductance, and poor scalability further constrain industrial prospects.[33-35]

To overcome these bottlenecks, researchers have advanced the field along three interconnected dimensions: theory, modulation, and experimental realization. Theoretical frameworks have evolved from simple tunneling models to multiscale transport descriptions that incorporate effects such as Fermi-level pinning and metal-induced gap states (MIGS).[36-39] First-principles simulations combining density functional theory (DFT) with non-equilibrium Green's function (NEGF) approaches now provide accurate predictions of structure-property relationships and shed light on

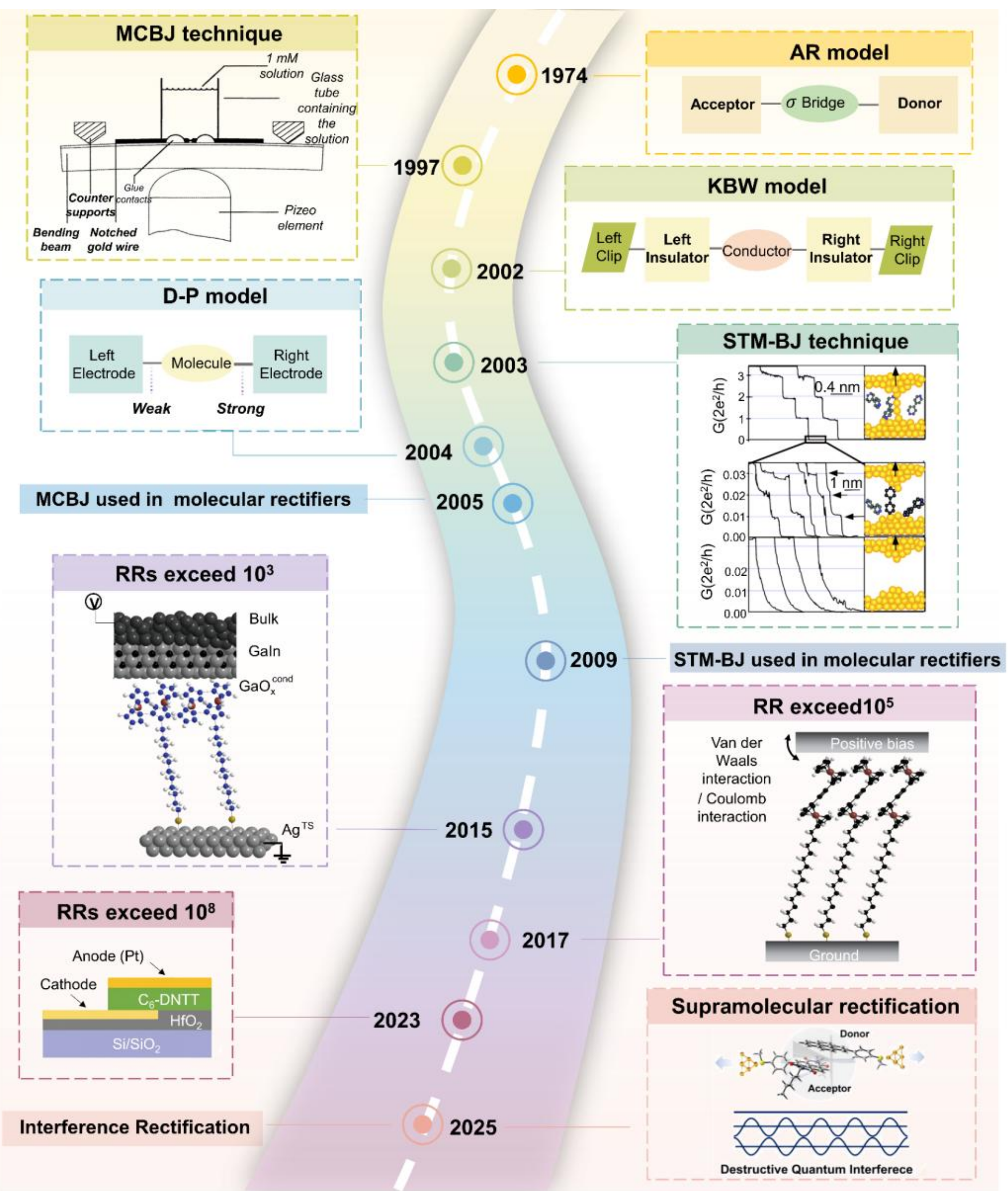

**Figure 1. Evolution roadmap and key milestones in the development of molecular rectifiers.** The roadmap illustrates the transition from theoretical concepts to experimental realization. Significant breakthroughs in rectification ratios (RR) exceeding $10^3$, $10^5$, and $10^8$ are highlighted, culminating in recent advancements in supramolecular architectures and quantum interference engineering. Reproduced with permission Ref. [48]. Copyright 1997, American Association for the Advancement of Science. Reproduced with permission Ref. [166]. Copyright 2003, American Association for the Advancement of Science. Reproduced with permission Ref. [80]. Copyright 2015, American Chemical Society. Reproduced with permission Ref. [22]. Copyright 2017, Springer Nature. Reproduced from Ref. [24] under the CC BY 4.0 license. Reproduced with permission Ref. [47]. Copyright 2025, American Chemical Society.

mechanisms such as quantum interference (QI) and interfacial charge transfer.[40-42] In parallel, the use of carbon-based electrodes, including graphene and carbon nanotubes, has markedly improved both interfacial stability and electrical performance. A wide range of external modulation strategies, such as structural engineering, and external-field modulation (e.g., optical, temperature, electrical, and chemical), have opened new avenues toward achieving high-performance rectifiers.[29, 43-49] Alongside these advances, device fabrication has progressed from early metal-vacuum-metal tunneling structures to highly controlled platforms based on scanning tunneling microscopy (STM),[50] mechanically controllable break junctions (MCBJ),[51] and SAM-based junctions.[52-55] More recently, emerging architectures such as cross-plane break junctions (XPBJ), electromigrated nanogap electrodes, and van der Waals contacts based on two-dimensional materials have significantly improved structural stability and junction reproducibility.[56, 57]

The study of molecular rectifiers has thus evolved from conceptual demonstrations to experimental synthesis and performance optimization.[58, 59] To provide a comprehensive overview of this field, this review is organized along the trajectory from mechanism to modulation and realization. The fundamental charge transport principles and classical rectification models are first introduced, followed by a discussion of modulation strategies that involve molecular design, interfacial engineering, quantum effects, and external-field responses. Advances in fabrication and characterization methods are then surveyed together with developments in first principles modeling. To illustrate the evolution of the field, **Fig. 1** provides a timeline spanning from 1974 to the present, highlighting theoretical milestones and experimental breakthroughs. Finally, major challenges and possible future directions are discussed, with an emphasis on the prospective role of molecular rectifiers within the post-Moore electronic landscape.

## 2 Theoretical Framework of Molecular Rectification

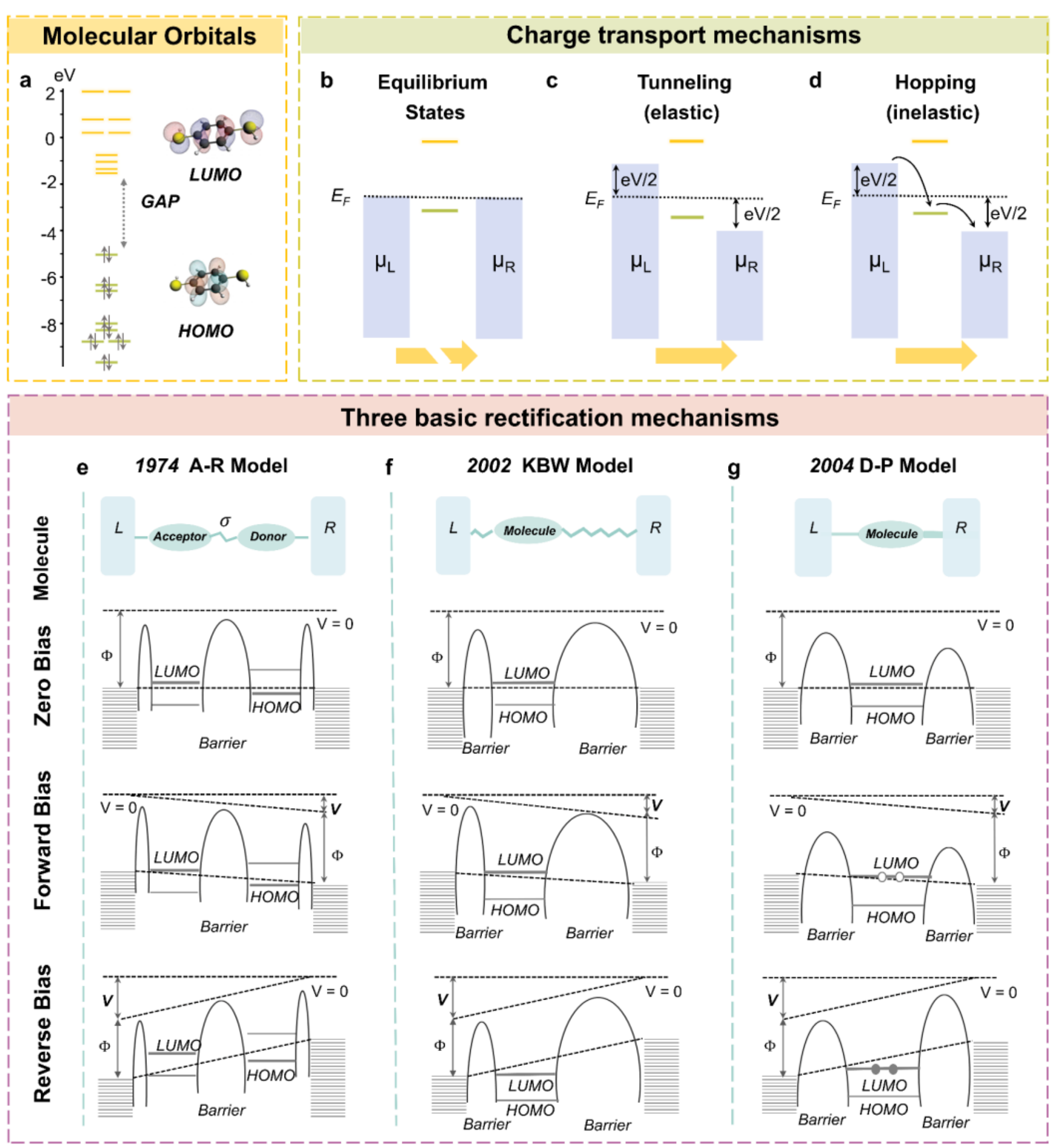

**Figure 2. Theoretical frameworks and charge transport mechanisms. (a)** Molecular energy spectrum and electronic states of the highest occupied molecular orbital (HOMO) and lowest unoccupied molecular orbital (LUMO) for a benzene dithiol molecule calculated by density functional theory (DFT) calculations. Green and yellow lines denote occupied and unoccupied orbitals, respectively. Reproduced with permission Ref. [19]. Copyright 2015, Royal Society of Chemistry. **(b-d)** Energy level diagrams illustrating **(b)** the equilibrium state, **(c)** the coherent quantum tunneling regime, and **(d)** the incoherent hopping-like charge transport regime. **(e-g)** Three rectification mechanisms and bias-dependent energy level evolutions: **(e)** Aviram-Ratner (AR) model based on a donor-σ-acceptor (D-σ-A) system with an insulating spacer, where forward bias facilitates resonant tunneling by aligning the acceptor LUMO and donor HOMO, whereas reverse bias inhibits current through an increased potential threshold; **(f)** the Kornilovitch-Bratkovsky-Williams (KBW) model with an asymmetric tunnelling barrier, where the majority of the applied bias drops across the longer insulating segment, inducing a bias-dependent onset for resonant tunneling; and **(g)** the Datta-Paulsson (DP) model, highlighting rectification driven by asymmetric molecule-electrode coupling strengths, which results in imbalanced orbital occupancy and asymmetric charging dynamics under bias. **(e-g)** Adapted with permission Ref. [79]. Copyright 2012, American Chemical Society.

The mechanisms underlying molecular rectification are inherently complex, arising from the interplay between molecular electronic structure, molecule-electrode interactions, and nonequilibrium charge transport under applied bias. A thorough understanding of these fundamental physical principles is therefore crucial for the rational design of high-performance molecular devices with large RRs and reliable operation. This section presents a systematic theoretical framework, from the electronic structure of molecular orbitals to the principal rectification mechanisms, laying the foundation for design of efficient molecular rectifiers.

### 2.1 Molecular orbitals

Charge transport at the molecular scale is governed by the principles of quantum mechanics.[60, 61] Owing to the extremely small dimensions of molecules, charge carriers are confined within a finite spatial region, resulting in a discrete energy level spectrum.[62] The spacing between adjacent molecular levels is determined primarily by the spatial dimensions of the system and the effective mass of the electrons, and can be qualitatively estimated using the particle-in-a-box model. In general, the smaller the molecular dimensions, the greater the degree of energy level discretization.

These discrete levels, defined as molecular orbitals (MOs), serve as the fundamental basis for describing the electronic states of a molecule.[63] Their spatial distribution is dictated by the electrostatic potential landscape generated by the nuclei within the molecule. Electron filling of these orbitals follows the Aufbau principle and the Pauli exclusion principle, with each orbital accommodating at most two electrons of opposite spin. By definition, the highest energy molecular orbital that is fully occupied by electrons is called the highest occupied molecular orbital (HOMO), which is analogous in physical significance to the valence band maximum in inorganic semiconductors. Conversely, the lowest unoccupied molecular orbital (LUMO) corresponds to the conduction band minimum. As an illustrative example, the energy spectrum of a benzene dithiol molecule in the gas phase is shown in **Fig. 2a**.[19] The right panel presents the isosurfaces of the HOMO and the LUMO wavefunctions, while the left panel shows the energy spectrum. The energy difference between the HOMO and LUMO is called the HOMO-LUMO gap, which is not only a key descriptor of the molecular electronic structure but also a critical parameter determining molecular charge transport properties.[64]

### 2.2 Tunneling Transport Mechanisms

When a molecule is linked to two electrodes via chemical bonding or physical adsorption, charge carriers traverse the junction through discrete molecular orbitals, while the electrodes act as reservoirs for electrons or holes.[65] Under equilibrium states, the chemical potentials of both electrodes are identical, resulting in zero net current, as illustrated in **Fig. 2b**. Upon applying a symmetric bias voltage, the chemical potentials of the left and right electrodes shift in opposite directions, forming a "bias window". If a molecular orbital lies within this window, carriers can be injected and extracted without an energy barrier, leading to substantial current flow. Conversely, if the orbital lies outside the window, transport occurs via quantum tunneling across an energy barrier, with efficiency governed by the alignment between molecular levels and electrode chemical potentials.[66]

Coupling between the molecular orbitals and the electrodes induces hybridization of electronic states.[67] This coupling strength not only dictates the position and broadening of the molecular energy levels, but also determines whether electrons can preserve their quantum phase during transport. Based on phase preservation, transport can be either coherent or incoherent.[68] In the coherent regime, the electron wavefunction retains phase coherence throughout the transport process. This typically occurs in short molecules with strong molecule-electrode coupling and can be described by the Landauer-Büttiker formalism, shown in **Fig. 2c**.[69, 70] In the incoherent regime, electrons dwell on the molecule for longer times and lose phase information due to molecular vibrations or environmental interactions, resulting in thermally activated, hopping-like conduction that can be modeled by theories such as the Marcus formalism (**Fig. 2d**).[71, 72]

Another independent classification dimension is based on the relative position of the molecular levels with respect to the bias window.[60] When the molecular level aligns with the electrode chemical potentials, resonant tunneling takes place, eliminating the energy barrier and greatly enhancing conductance. When the level lies outside the bias window, off-resonant tunneling occurs via virtual states, and the current decays exponentially with increasing energy mismatch. These two classification schemes are orthogonal. Coherent transport may occur under resonant conditions, giving rise to elastic resonant tunneling, or under off-resonant conditions as elastic off-resonant tunneling. Likewise, incoherent transport can proceed under resonant conditions via thermally activated sequential tunneling or under off-resonant conditions via thermally activated off-resonant tunneling.

### 2.3 Landauer-Büttiker Formalism

In this review, we primarily adopt the Landauer framework to describe coherent tunneling in molecular junctions. Within this picture, the junction is treated as a phase-coherent conductor, and all inelastic interactions are neglected. The current is then expressed as: [69, 70]

$$I = \frac{2e}{h}\int_{-\infty}^{\infty}(T(E)(f_L(E) - f_R(E)))dE \tag{1}$$

where $e$ is the electron charge, $h$ is Planck's constant, and $f_{L,R}(E)$ are the Fermi-Dirac distribution functions of the left and right electrodes. $T(E)$ is the transmission probability for electrons of energy $E$. Within the non-equilibrium Green's function (NEGF) formulism, the transmission can be expressed as:[73]

$$T(E,V) = tr(\Gamma_L G^R \Gamma_R G^A) \tag{2}$$

where $G^R$ and $G^A$ are the retarded and advanced Green functions of the molecular region, given by $G^R(E) = \frac{1}{E-H-\sum_L(E)-\sum_R(E)}$ and $G^A(E) = \frac{1}{E-H-\sum_L(E)-\sum_R(E)-i\delta}$. Here, $H$ is the Hamiltonian of the central region, while $\sum_L(E)$ and $\sum_R(E)$ are the self-energies associated with the left and right electrodes. The coupling matrices $\Gamma_{L/R}(E) = i\left[\sum_{L/R}(E) - \sum_{L/R}^{\dagger}(E)\right]$ quantify the orbital overlap between molecule and electrodes.

For a single-level model considering only a spin-degenerate orbital of energy $E_0$, the transmission function reduces to a Lorentz form:[74, 75]

$$T(E) = \frac{4\Gamma_L\Gamma_R}{[E - E_0]^2 + [\Gamma_L + \Gamma_R]^2} \leq 1 \tag{3}$$

This expression reveals two key principles. First, transmission reaches unity under resonant conditions $(E = E_0)$ when the couplings are symmetric $(\Gamma_L = \Gamma_R)$. Second, transmission increases with coupling strength, but excessive coupling leads to pronounced level broadening, diminishing the localization of molecular states and potentially degrading device functionality.[76]

The Landauer-Büttiker formalism therefore provides not only a direct mathematical link between macroscopic current and microscopic transport processes, but also a unified theoretical framework for understanding how molecular energy levels, interfacial coupling, applied bias, and temperature collectively determine the performance of molecular electronic devices.

### 2.4 Rectification Mechanisms

Molecular rectification in single-molecule devices is fundamentally governed by distinct mechanisms that arise from structural asymmetry, potential landscape variations, and interfacial coupling differences. Three prototypical theoretical models, the AR model, the Kornilovitch-Bratkovsky-Williams (KBW) model, and the Datta-Paulsson (DP) model, are three prototypical theoretical frameworks for explaining the rectification behavior of single-molecule diodes, as schematically illustrated in **Figs. 2e-g**.[12, 77-79]

The AR model, first proposed by Aviram and Ratner, embodies the conceptual foundation of molecular diodes through a D-σ-A architecture (**Fig. 2e**).[12] Here, an insulating σ-bridge spatially separates donor and acceptor units, preserving their individual electronic identities and minimizing direct coupling. Charge transport proceeds via sequential tunneling: electrons transfer from the left electrode to the acceptor, then through the donor, before reaching the right electrode. At equilibrium, the donor's HOMO lies just below the electrode Fermi level, while the acceptor's LUMO is positioned slightly above it. Under forward bias, this energy gap narrows, facilitating resonant tunneling and reducing the conduction threshold. Conversely, reverse bias requires significantly higher voltages to align orbitals, resulting in pronounced current asymmetry. While elegant in theory, the synthetic complexity and limited tunability of the AR framework restrict its practical implementation. To address these limitations, derivative systems such as donor-π-acceptor (D-π-A) systems and donor-acceptor block copolymers have been developed, retaining the core rectification mechanism while enhancing chemical accessibility and design flexibility.[80-83]

The KBW model offers a complementary perspective by attributing rectification to an asymmetric tunneling barrier interacting with a single molecular orbital, thereby dispensing with the need for distinct donor and acceptor moieties (**Fig. 2f**).[77] Since most of the applied bias drops across the longer insulating segment, the molecular orbital responds asymmetrically under forward and reverse biases, thereby leading to a bias-dependent onset voltage for resonant tunneling. By adjusting the length of the insulating tails, the RRs can also be effectively tuned.[24, 84] Compared to the AR model, the KBW framework relaxes stringent energy level alignment requirements, thereby enhancing synthetic feasibility and experimental realization.

Distinct from the intramolecular asymmetry-based mechanisms, the DP model emphasizes the role of asymmetric molecule-electrode coupling in inducing rectification, even when the molecule itself is spatially symmetric (**Fig. 2g**).[78] Unequal coupling strengths to the electrodes result in imbalanced charging and discharging dynamics of molecular orbitals under bias, yielding direction-dependent occupancy and consequently asymmetric current flow. Although engineering precise interfacial asymmetry remains experimentally challenging, this mechanism is frequently observed in molecular junction measurements and plays a crucial role in device performance.[85]

Collectively, these three models capture essential mechanisms of molecular rectification from three distinct perspectives. In practical device engineering, effective rectification typically requires a synergistic integration of multiple mechanisms to balance structural tunability with transport performance. These three theoretical frameworks have not only laid the foundational groundwork for the field but also continue to serve as essential guiding principles for the design and optimization of molecular diodes.

# 3 Modulation Strategies

The theoretical framework of molecular rectifiers indicates that their rectification performance arises from the interplay of multiple factors. Building on rectification models and molecular transport theory, researchers have developed diverse modulation strategies, including molecular structural, interfacial, quantum interference, and external field engineering.[8, 86, 87] This section classifies these representative approaches to molecular rectification, offering methodological guidance for the design of high-performance molecular rectifiers.

## 3.1 Molecular Structural Engineering

Modulating intrinsic molecular structures is among the most fundamental strategies for enhancing rectification behavior.[88] As discussed above, both the AR and KBW models rely on molecular design to achieve rectification. However, molecular-structure engineering not only encompasses traditional molecular backbone architectures, but also emphasizes precise control over torsional angles of bridging units, the length of the molecular backbone, the configuration of side chains, and the conjugation mode.

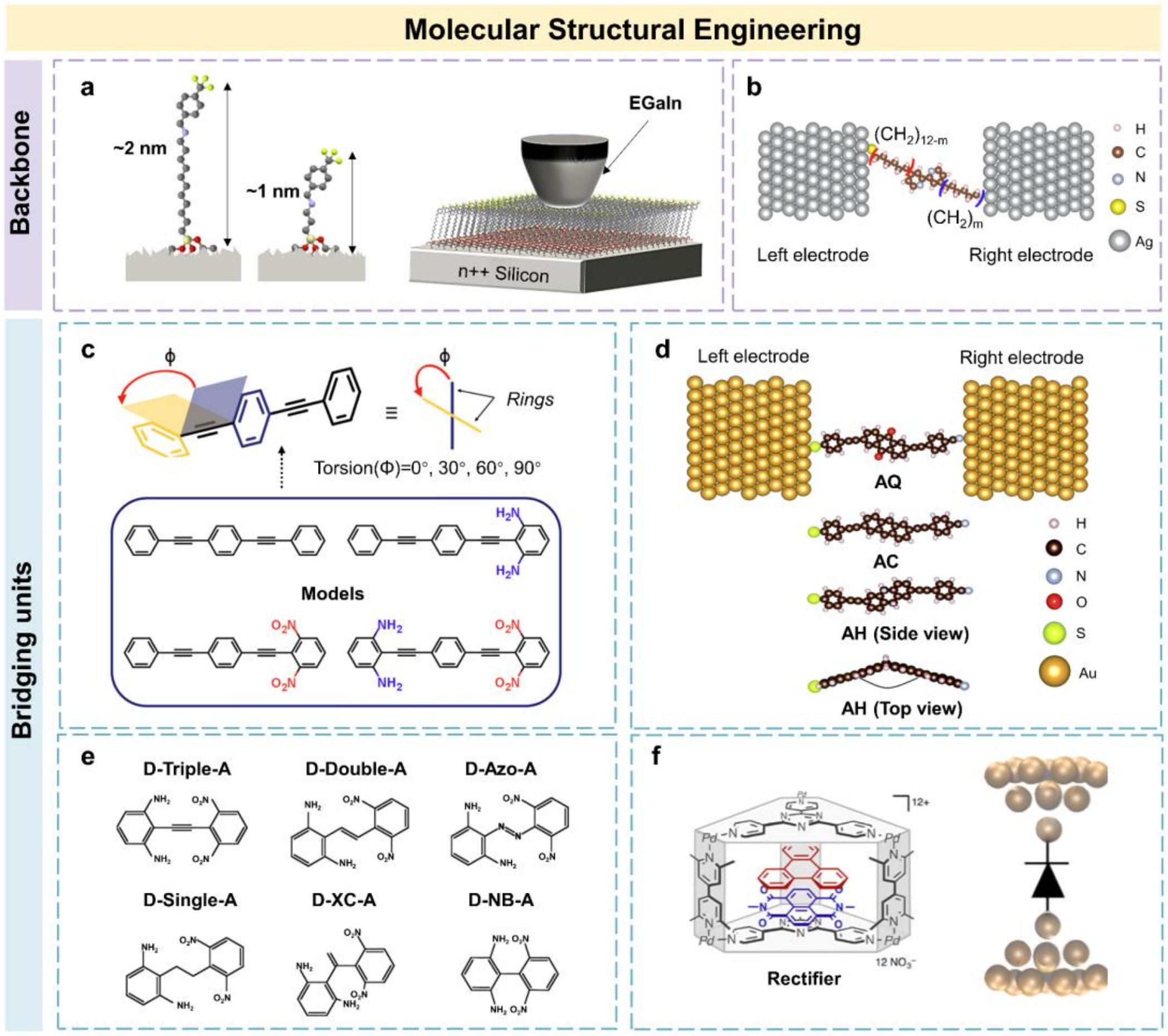

**Figure 3. Molecular structural engineering strategies.** (a) Modulation of RRs via chain length in alkylated silane self-assembled monolayer (SAM) rectifiers. Reproduced from Ref. [92] under the CC BY 4.0 license. (b) Control of rectification direction by positioning the bipyridyl unit (*m* value) within an alkanethiolate backbone. Reproduced with permission Ref. [89]. Copyright 2017, American Chemical Society. (c) Bridging unit engineering through torsional angle control and chemical substitution in phenylene-based D-B-A systems. Reproduced with permission Ref. [97]. Copyright 2018, American Chemical Society. (d) Device models for thiolated arylethynylene diodes utilizing cross-conjugated (AQ), linearly conjugated (AC), and broken-conjugated (AH) central bridges to modulate rectification. Reproduced with permission Ref. [99]. Copyright 2017, Royal Society of Chemistry. (e) Investigation of D-B-A systems with six diverse bridge architectures (linear, cross-conjugated, and saturated). Reproduced with permission Ref. [111]. Copyright 2023, American Chemical Society. (f) Supramolecular rectification in a self-assembled coordination cage containing hetero $\pi$-stacked aromatic pairs (naphthalenediimide and triphenylene). Reproduced with permission Ref. [112]. Copyright 2015, American Chemical Society.

Introducing asymmetry through backbone engineering allows effective modulation of RRs. At the macroscopic level, this asymmetry manifests as an inhomogeneous potential distribution across the device, while at the microscopic level it corresponds to spatial variations in molecular orbital distributions. Typically, altering molecular type or length represents the most straightforward and widely employed strategy. This approach is broadly applicable across various molecular rectifier models. Length modulation is particularly relevant in KBW models and long alkyl-chain systems, where the pronounced "tail" structure renders rectification highly sensitive to molecular length.[89-91] However, the effect of chain length varies among different systems. In alkylated silane SAMs, increasing molecular length reduces the RR, which correlates positively with molecular dipole strength (**Fig. 3a**).[92] In bipyridyl-embedded alkanethiolate molecules, rectification occurs under positive bias for $1 \leq m \leq 3$ carbon atoms on the bipyridyl's right side, whereas for $5 \leq m \leq 11$, the rectification direction reverses (**Fig. 3b**).[93] This behavior originates from the monotonic evolution of strongly localized FMO under applied bias. Fine-tuning the position of the bipyridyl unit along the alkanethiolate backbone allows precise control of the orbital spatial distribution, thereby dictating the rectification direction. This variation in chain-length effects reflects differences in backbone architecture and dipole localization. In alkylated silane SAMs, dipoles become increasingly delocalized as the chain lengthens, whereas in bipyridyl-embedded systems, the rigid bipyridyl group directly determines the center of orbital localization.

In addition to altering the intrinsic molecule architecture, strategies such as elemental doping, functionalization, and side-chain modification have also proven effective in tuning rectification (**Fig. 3c**).[56, 94-97] Doping or functional groups modulate the molecular orbital landscape, whereas side chains can act as an internal "gate", adjusting the alignment of molecular orbitals relative to the electrode Fermi level. Despite substantial progress in both theory and experiment, observed RRs frequently fall short of model predictions, and measured rectification direction occasionally diverges from theoretical expectations. Moreover, for large or multi-unit molecules, synthetic complexity and molecular heterogeneity remain critical barriers, underscoring the need for precise synthetic control over backbone architecture and functional group positioning to bridge the gap between molecular design and device performance.

The geometry and conjugation characteristics of bridging units also play a pivotal role in modulating the rectification performance of molecular junctions, as clearly demonstrated by the AR model. Bridging units directly govern the electronic coupling between the donor and acceptor, whereas the essence of rectification lies in achieving effective electronic decoupling between these two moieties. Tuning the length of the molecular bridge represents an efficient strategy to weaken electronic coupling, thereby enhancing the RR, that has been confirmed in pyrene-benzene molecular junctions.[98] In addition, torsional distortions of the bridging units can partially separate the electronic states of the donor and acceptor, further promoting rectification (**Fig. 3d**).[99, 100] A recent systematic study compared different types of bridging structures in terms of their influence on transport behavior and rectification performance, revealing their intrinsic characteristics.[101] As shown in **Fig. 3e**, the investigated bridging architectures are categorized into three main types: linear conjugated bridges (triple bonds, double bonds, and azo bridges), cross-conjugated bridges (XC), and saturated bridges (single bonds and no bond (NB) bridges). The results indicate that cross-conjugated bridges achieve the highest RR, reaching 2.32 under a bias of 0.15 V. This is primarily attributed to the absence of a direct conjugation pathway between the donor and acceptor in XC systems, which confines the HOMO strictly to the donor and the LUMO strictly to the acceptor, thereby inducing pronounced orbital spatial separation. However, this configuration exhibits extremely low conductance, highlighting the inherent trade-off between RR and electrical conductivity in bridge design. Moreover, owing to the synthetic challenges associated with these structures, most studies remain at the DFT level, and the performance of such junctions falls significantly short of bulk devices.

Beyond covalent connectivity, supramolecular assemblies based on π-π interactions, such as D-π-π-A systems, also exhibit notable rectification behavior (**Fig. 3f**).[102] Recent advances in scanning tunneling microscopy (STM) techniques have enabled precise control over the π-π stacking geometry in single-molecule supramolecular junctions.[103] Nevertheless, the high environmental sensitivity of noncovalent interactions endows such supramolecular rectifiers with stimulus-responsive characteristics, posing additional challenges for further enhancement of the RRs.

## 3.2 Interfacial Engineering

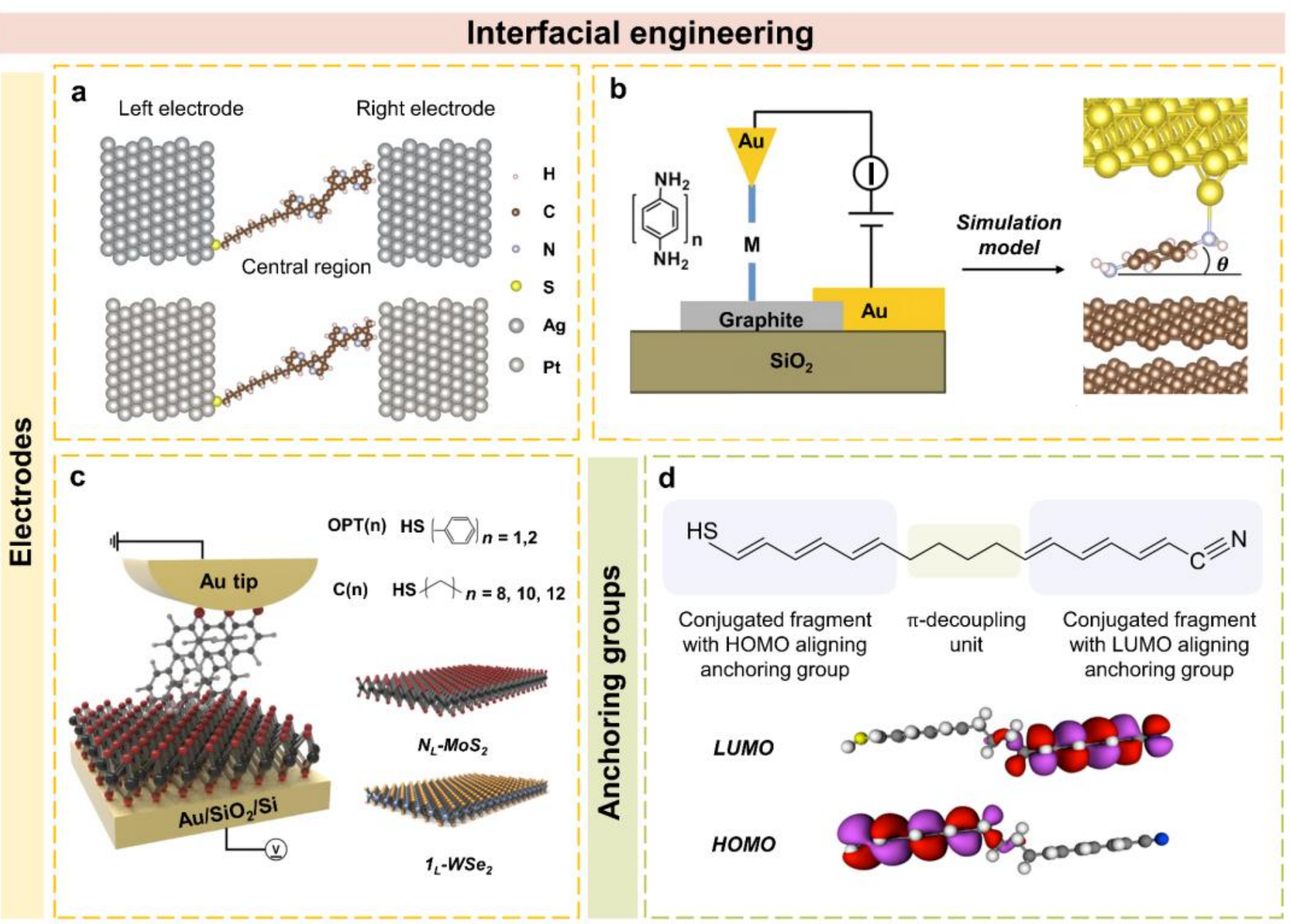

**Figure 4. Interfacial and electrode engineering strategies.** (a) Enhancement of RRs in $SC_{11}$BIPY–C≡C–BIPY diodes by replacing Ag with Pt electrodes. Reproduced with permission Ref. [108]. Copyright 2021, American Chemical Society. (b) Asymmetric molecular junction and its simulation model comprising a graphite substrate and an Au tip. Reproduced with permission Ref. [112]. Copyright 2014, National Academy of Sciences. (c) Tailored-diode behavior in molecular heterojunctions using 2D semiconductors ($MoS_2$ or $WSe_2$) as electrodes. Reproduced from Ref. [118] under the CC BY 4.0 license. (d) Rectification mechanism based on Fermi-level pinning using asymmetric anchoring groups and $\pi$-decoupling units. Reproduced with permission Ref. [128]. Copyright 2015, American Chemical Society.

The Landauer formalism and the DP model both reveal, at the theoretical level, that rectification is critically determined by the strength of molecule-electrode coupling. This coupling can be effectively modulated by altering electrode materials, optimizing interfacial contact geometries, and employing tailored anchoring groups.[88, 104-106] Such approaches provide powerful means to tune interfacial coupling, thereby inducing or enhancing rectification effects.

The choice of electrode materials plays a decisive role in determining the injection barriers and the electronic coupling strength at the molecule-electrode interface.[105] Metals such as Au, Ag, Pt, Pd, and Cu have been widely used as electrodes in molecular rectifiers, and their varying work functions create distinct charge injection barriers that strongly influence rectification.[107] Theoretical calculations by Zhang et al. showed that replacing Ag with Pt in a HOMO-dominated molecular system can enhance the RR by orders of magnitude, as high work function metals reduce hole injection barriers and facilitate efficient carrier transport (**Fig. 4a**)[108]. Other complementary studies suggest that the effect primarily stems from the molecule-electrode coupling rather than energy-level alignment, with the density of states (DOS) at the Fermi level playing a key role.[109-111] However, in either case, electrode selection inevitably modulates device rectification. Compared to metallic electrodes, carbon-based electrodes offer unique advantages due to their excellent interfacial compatibility and tunability (**Fig. 4b**).[112] Coupling with molecules primarily relies on covalent bonding or π-π stacking, which are typically weaker than metallic chemical bonds. This "weak coupling" preserves the intrinsic molecular states and allows precise control over energy-level alignment. Through doping, gate modulation, or interface functionalization, transport characteristics can be flexibly adjusted, providing greater freedom to optimize molecular rectification.[113]

Beyond molecular asymmetry, electrode heterogeneity represents another effective route for enhancing RRs.[114-117] Asymmetric electrode configurations can vary contact areas and interface spacing, working together with intrinsic electrode properties to shape device performance. In a landmark study, Nijhuis and colleagues achieved RRs exceeding $10^5$ in a Fc–C≡C–Fc (Fc: ferrocenyl) system by constructing asymmetric electrode configurations.[24] As illustrated in **Fig. 1**, one end of the molecule formed stable chemical bonds with a metal electrode, whereas the opposite end relied on van der Waals interactions with an EGaIn electrode. Under reverse bias, Coulomb interactions between the molecule and EGaIn strengthened the tunneling process, whereas in forward bias, van der Waals interactions dominated transport and generated pronounced rectification. More recently, two-dimensional (2D) semiconductors such as $MoS_2$ and $WSe_2$ have been incorporated as heterogeneous electrodes in organic molecule/metal interfaces, further enriching the strategies for electrode engineering (**Fig. 4c**). The RRs can reach as high as $1.83 \times 10^4$ in such systems, with the performance being highly tunable through variations in the type and number of 2D layers.[118] Overall, molecular rectifiers leveraging electrode and interfacial asymmetry are emerging as a mainstream research direction, with device performance approaching or even rivaling conventional inorganic diodes.[58]

Anchoring groups act as essential bridges between molecules and electrodes, where their conductivity and chemical nature are critical for molecular rectifiers.[119-121] Since Kushmerick et al. first demonstrated in 2004 that tuning anchoring groups can modulate rectification on demand, this approach has become a central strategy in interface engineering.[122] Unlike molecular backbone optimization, anchoring groups act at the injection bottleneck, and enable precise control of transport while preserving the intrinsic molecular electronic states.

Depending on bonding mechanism and electronic properties, anchoring groups can be classified into two types. Covalent groups, such as –SH and Au–C, enhance interfacial electronic coupling through strong chemical bonding.[123, 124] Donor or acceptor groups, including –CN, $–NH_2$, and $–NO_2$, modulate energy level alignment via electron transfer or lone-pair donation.[125-127] Anchoring groups with donor or acceptor characteristics are particularly impactful in this regard. By exploiting Fermi-level pinning, they can couple the HOMO and LUMO selectively to different electrodes. A representative example is provided by Van Dyck and Ratner, who introduced asymmetric donor and acceptor groups at opposite molecular termini and achieved RRs exceeding two orders of magnitude (**Fig. 4d**).[128] Under applied bias, the pinning effect drives asymmetric shifts of molecular orbitals, thereby altering tunneling probabilities in opposite bias directions and generating strong rectification behavior. Building on such design principles, recent advances in fullerene-based molecular diodes have pushed the RRs above $10^4$, underscoring the potential of anchoring-group design in high-performance devices.[129]

Despite these advances, some anchoring groups suffer from multiple binding configurations and weak stability, which compromise device reproducibility and lifetime.[130, 131] Coordinated efforts between theory and experiment to optimize bonding configurations, together with in situ characterization techniques, are expected to drive this strategy toward controlled, reliable, and scalable device implementation.

### 3.3 Quantum Interference Engineering

Quantum interference effects (QIEs) provide an alternative pathway to molecular rectification beyond conventional energy-level alignment.[47, 132] They originate from the phase superposition of electron wavefunctions along multiple transport channels, manifesting as constructive interference (CQI) that enhances current, or destructive interference (DQI) that suppresses it (**Fig. 5a**) [17, 133] Although highly sensitive to molecular structure and symmetry, QIEs have evolved from a physical phenomenon into a rational design principle.[134] Current studies on QIEs in molecular rectifiers largely fall into two categories: modulation through intramolecular configuration and intermolecular interactions.

Intramolecular configuration control is the most established strategy for interference-driven rectification, with aromatic systems serving as the primary platform. Early studies showed that asymmetric para-meta connected diphenyl-oligoenes intrinsically display direction-dependent current due to bias-sensitive coupling at the two terminals.[135] Subsequent work revealed that cross-conjugated motifs can also introduce interference channels, enhancing RRs without compromising conductance.[136] Among these, para- and meta-substituted benzenes remain the most intensively investigated.[137] Meta linkages generate DQI because of multiple phase-incoherent transport pathways, strongly suppressing current, whereas para linkages exhibit CQI through coherent path superposition, thereby promoting efficient charge transmission. The central challenge, however, lies in realizing dynamic switching between para and meta configurations. A recent breakthrough came from Guo and co-workers, who employed an electric-field-catalyzed Fries rearrangement to achieve reversible para-to-meta conversion (**Fig. 5b**).[48] Under forward bias, the molecule preserves the para structure and exhibits CQI. Under reverse bias, it is transformed into the meta structure, introducing DQI and switching the interference channel with bias polarity. Assisted by Lewis acid catalysis, this molecular rectifier delivered a RR exceeding 5,000 at 1.0 V, together with excellent cycling stability and reversibility. This work provides a substantive step forward in applying quantum interference to single-molecule rectification.

Beyond molecular backbones, intermolecular interactions can also trigger and regulate QIEs.[49, 134, 138] Noncovalent forces such as $\pi$–$\pi$ stacking not only stabilize supramolecular assemblies but also enable phase coupling between stacked orbitals, periodically toggling between CQI and DQI to generate rectification. A representative example was recently reported by Xu et al., who designed a D–$\pi$–$\pi$–A single-molecule rectifier assembled from pyrene (Py) and naphthalenediimide (NDI) through $\pi$–$\pi$ interactions (**Fig. 5c**).[49] The device exhibited pronounced bias-dependent asymmetry in current response. More importantly, mechanical stretching induced a reversible structural reconfiguration, amplifying energy-level asymmetry and transport-pathway differences, thereby achieving a significant enhancement of the RR. This study provides the first experimental validation that quantum interference can be modulated via intermolecular interactions, offering a new route to flexible and high-performance molecular rectifiers.

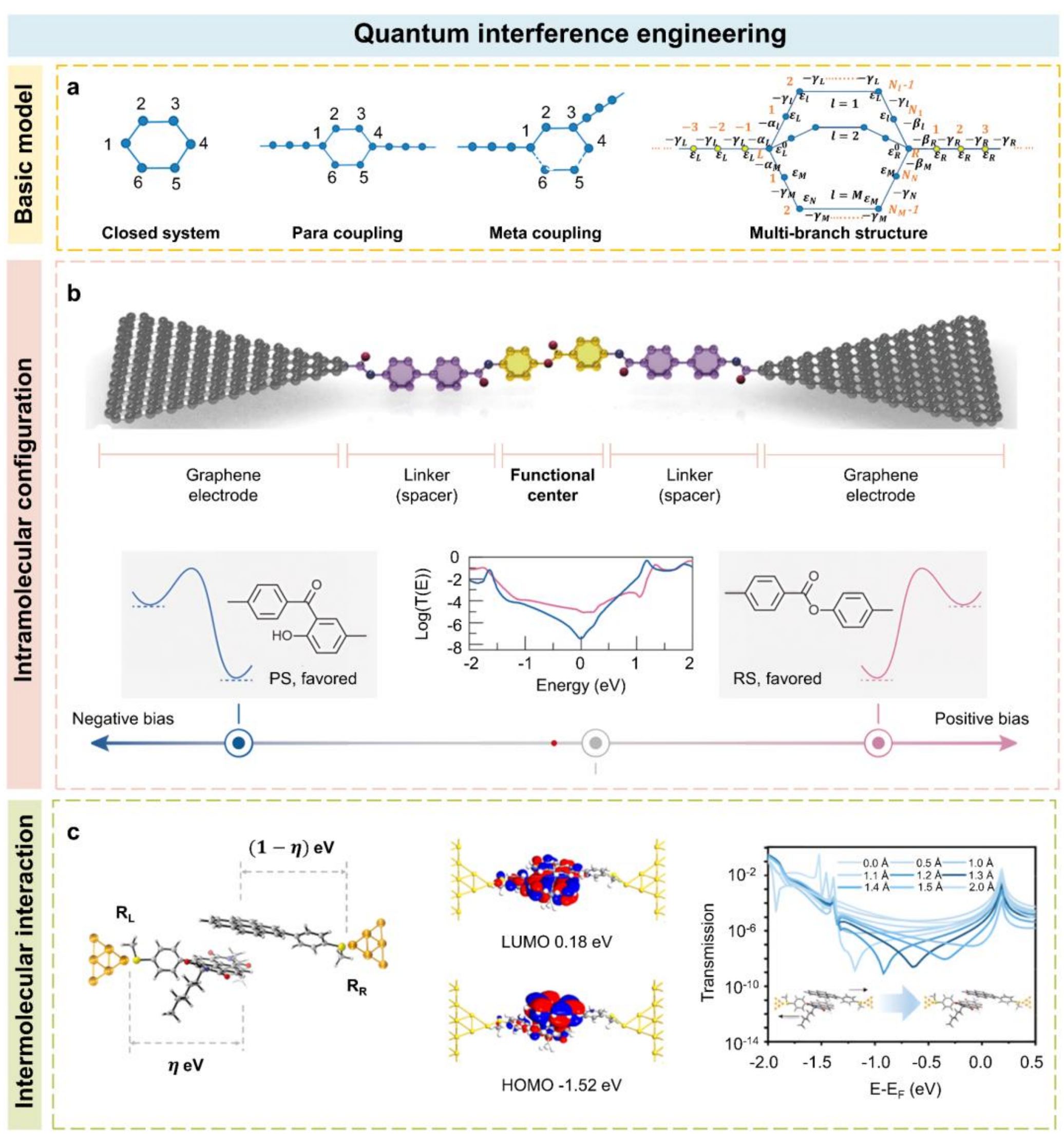

**Figure 5. Quantum interference (QI) engineering strategies. (a)** Theoretical models of QI in molecular junctions, including para-coupling (constructive QI), meta-coupling (destructive QI), and a generalized multi-branch structure; the transmission $T(E)$ is determined by site energies $\varepsilon$ and hopping elements $\gamma$, where the connectivity dictates the interference patterns. Reproduced with permission Ref. [19]. Copyright 2015, Royal Society of Chemistry. **(b)** Intramolecular configuration: schematic of a single-molecule junction with a phenyl benzoate functional center. The transport symmetry is broken by the asymmetric energy profiles under opposing electric field directions, as evidenced by the shifted transmission spectra for the rectified (RS) and non-rectified (PS) states. Reproduced with permission Ref. [50]. Copyright 2025, American Chemical Society. **(c)** Intermolecular interaction: supramolecular junctions with asymmetric electrostatic potentials, showing optimized frontier molecular orbital (FMO) distributions and transmission spectrums at varying electrode separations. Reproduced with permission Ref. [51]. Copyright 2025, American Chemical Society.

Nevertheless, progress in this direction remains limited, largely constrained by the need for precise structural control. With advances in supramolecular chemistry and dynamic assembly strategies, however, the potential of this approach is beginning to unfold.

## 3.4 External Field Modulation

External field modulation has emerged as an important auxiliary strategy for dynamically tuning molecular rectification. By applying physical or chemical perturbations, molecular or device systems can undergo structural reorganization, energy level shifts, or charge redistribution. These processes enable precise modulation of electronic rectification characteristics. Typical external fields include electric fields, temperature fields, optical fields, and chemical environment modulation.[18, 27, 46, 139, 140]

### *3.4.1 Electric Field Modulation*

Electric field modulation represents the most established strategy among external stimuli, relying on gate electrodes to tune molecular orbital energies or interfacial chemical potentials. Essentially, this process manifests as the interaction between molecules and an applied electric field, specifically the modulation of molecular orbital energies by the Stark effect.[141, 142] Building on this core mechanism, electric field modulation has evolved into two complementary paradigms. The first is the solid-state gate, an early strategy that relies on energy-level alignment. In contrast, the more recently developed liquid gate exploits the adsorption and desorption of ions at the electrode surface to indirectly modulate the electrode potential, thereby influencing molecular charge transport.

For instance, Perrin et al. demonstrated a gate-tunable single-molecule rectifier with a RR as high as 600 (**Fig. 6a**).[143] The key advance was the transition between negative differential conductance and rectification, supported by combined theoretical and experimental evidence. A negative gate bias shifts the orbital energy closer to the Fermi level. This reduces the injection barrier and thereby enhances rectification. However, the performance of solid-state gates is strongly dependent on device geometry and requires precise control of molecular conformation. Dual-gate configurations can improve modulation precision, yet excessively strong fields may distort molecular structures, limiting their practical utility.

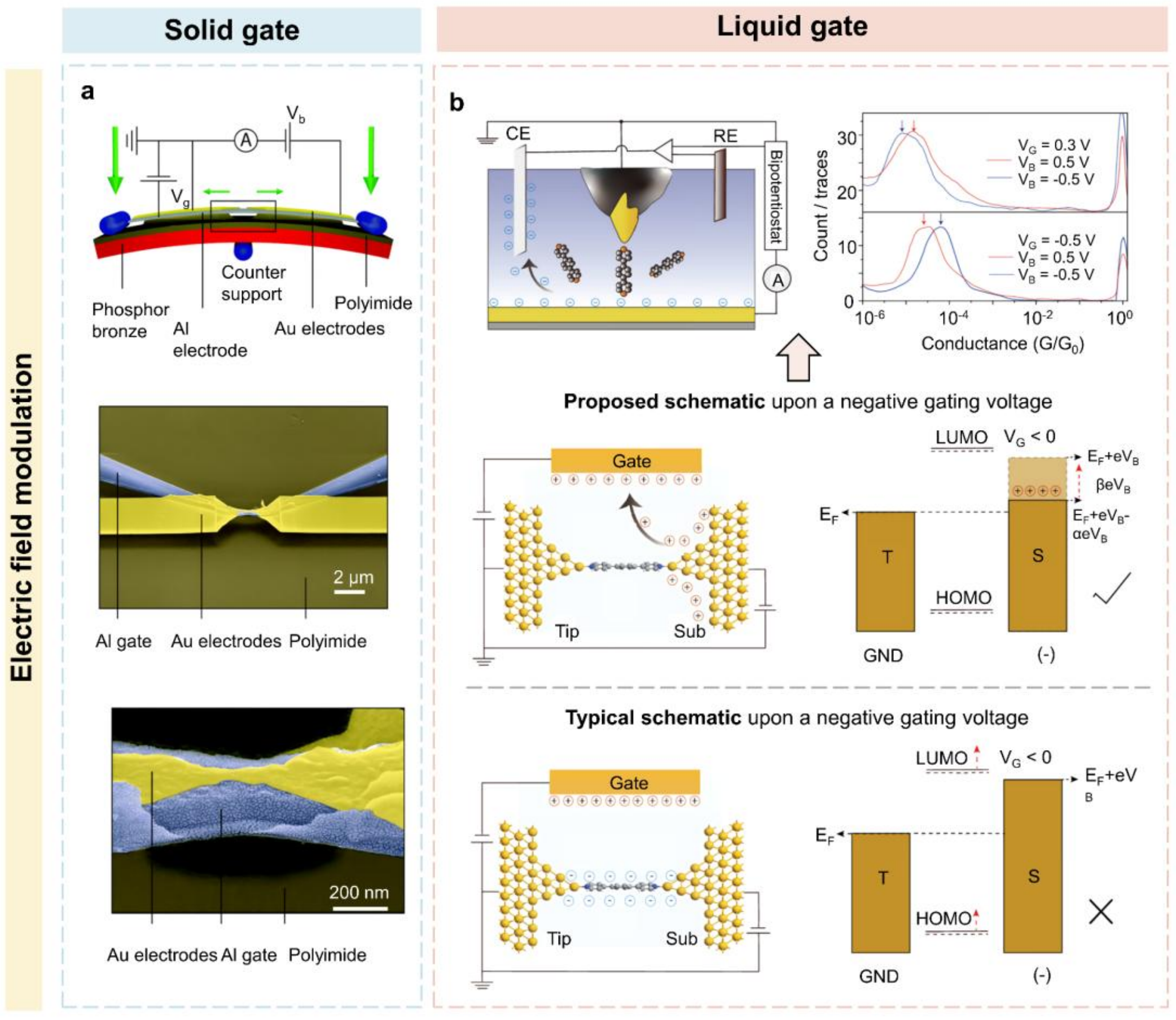

**Figure 6. Electric field modulation strategies. (a)** Solid gate modulation: schematic and scanning electron microscope (SEM) images of a gated mechanically controllable break junction (MCBJ) device. Reproduced with permission Ref. [143]. Copyright 2016, Royal Society of Chemistry. **(b)** Liquid gate modulation: schematic of an ionic liquid-gated single-molecule junction using the scanning tunneling microscope break Junction (STM-BJ) technique with a four-electrode configuration. The corresponding conductance histograms and energy-level diagrams illustrate the proposed ionic gating mechanism, where gating-induced ion redistribution modulates the substrate Fermi level ($E_F$) and the HOMO-LUMO alignment compared to the typical schematic. Reproduced with permission Ref. [31]. Copyright 2024, American Chemical Society.

Liquid gates, in contrast, circumvent these constraints. They provide efficient modulation at low operating voltages below 1 V and are insensitive to structural variations of the device. As illustrated in **Fig. 6b**, Wang and colleagues demonstrated in a symmetric electrode-molecule configuration that liquid gating can even invert the rectification direction in situ. [29] Their study revealed that, within ionic liquids, gate bias no longer directly shifts orbital energies but instead regulates the dynamic reconstruction of the interfacial electric double layer through the adsorption-desorption of surface-bound ions. While the applied bias governs the initial ion adsorption, the gate voltage primarily drives ion release, resulting in precise control of rectification direction.

Together, these advances highlight the versatility of electric field modulation in enhancing rectification performance and enabling reversible and low-power operation. Nevertheless, intrinsic limitations remain. Solid-state gating is constrained by geometric precision and molecular stability, while liquid gating, despite its adaptability, depends on complex ionic environments whose long-term stability and integration potential are yet to be established. Overcoming these challenges will be critical for realizing large-scale integration and practical applications.

### ***3.4.2 Temperature Modulation***

Temperature plays a crucial role in elucidating the competition between tunneling and thermally activated mechanisms in molecular rectifiers.[144] For small-scale molecular devices (< 4 nm), charge transport is predominantly governed by coherent tunneling and shows minimal temperature dependence.[145] In contrast, larger systems exceeding 4 nm are dominated by incoherent, thermally activated transport, which can be described within the Marcus framework.[146, 147] In classical Marcus theory, increasing temperature generally accelerates thermally activated charge transfer, following the exponential dependence of the Arrhenius equation on activation energy and temperature. However, recent studies have increasingly challenged this conventional view. Many molecular junctions exhibit behaviors that cannot be fully captured by a single theoretical model. Long-range tunneling has been observed in protein-based systems spanning 7 nm, whereas thermal activation dominates transport in molecular rectifiers shorter than 3 nm.[148-151] In reality, the temperature dependence of charge transport in molecular junctions is strongly modulated by both the magnitude and polarity of the applied bias, giving rise to highly complex response patterns.[152] This rich dependence provides a novel strategy for the directional optimization of molecular rectifiers.

Self-assembled monolayer devices provide a prototypical platform for temperature control, as their dimensions enable the manifestation of thermally dependent rectification. The underlying mechanism involves a dynamic switching between Marcus and inverted Marcus regions induced by the bias polarity. In high-performance aromatic SAM-based molecular rectifiers, as depicted in **Fig. 7a**, forward bias drives charge transport predominantly through thermally activated tunneling, whereas reverse bias exhibits a negative temperature coefficient.[153] This behavior arises from the enhanced conformational freedom at elevated temperatures. The charge state of polarizable terminal groups evolves with temperature, and the bias dependence of the activation energy leads to a cooperative effect between temperature and voltage on the number of conductive molecules. Similar mechanisms have been confirmed in multiple systems. For example, in benzotetrathiafulvalene (BTTF)-based junctions, temperature modulation can enhance the RRs by up to thirtyfold (**Fig. 7b**).[151] In certain cases, thermal fields even enable complete reversal of rectification direction (**Fig.**

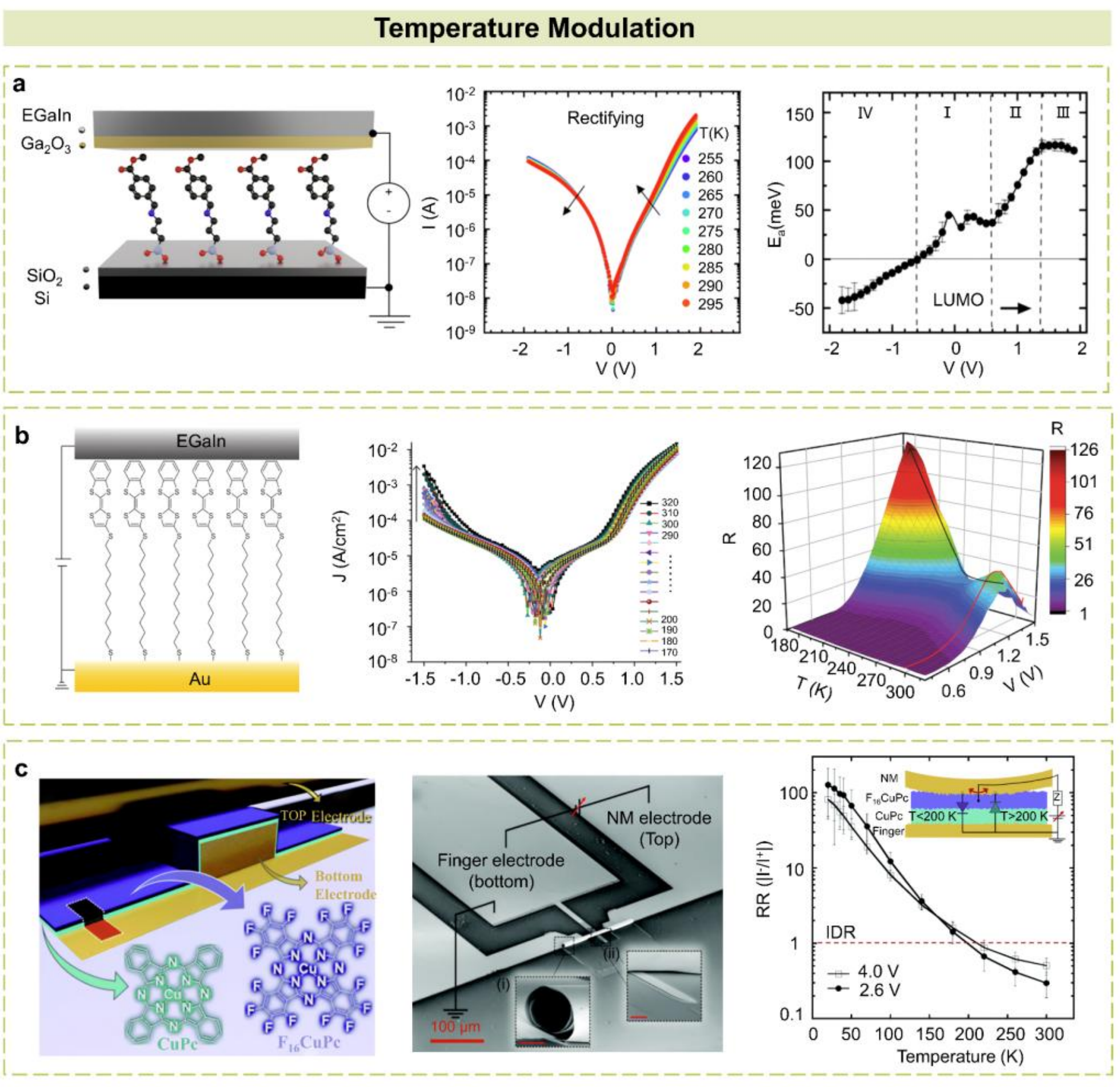

**Figure 7. Temperature modulation strategies. (a)** Device structure and current-voltage (I-V) characteristics of a CMPTM-based rectifying SAM, with Arrhenius plots illustrating temperature-dependent transport across four regimes: (I) off-resonant, (II) near-resonant, (III) resonant, and (IV) off-resonant. Reproduced with permission Ref. [153]. Copyright 2024, Royal Society of Chemistry. **(b)** The rectifier structure, temperature-dependent current density-voltage (J-V) curves, and 3D surface plot of the RR evolution for Au-S$(CH_2)_{11}$S-BTTF//$GaO_x$/EGaIn junctions. Reproduced from Ref. [151] under the CC BY 4.0 license. **(c)** Molecular diode based on CuPc and $F_{16}$CuPc heterostructures, featuring SEM images and the average RRs with statistical deviations as a function of temperature. The equivalent circuit illustrates the temperature-dependent device impedance (Z) via a thermal switch model. Reproduced with permission Ref. [46]. Copyright 2020, Royal Society of Chemistry.

**7c)**.[46] Collectively, these findings demonstrate that precise control over charge transport through different hopping regions, including the inverted Marcus regime, can substantially optimize rectification performance and guide the design of other thermally responsive platforms.

It is important to recognize the inherent limitations of temperature control. Elevated temperatures can induce interface reconstruction or molecular conformational changes, leading to irreversible degradation of device performance and posing challenges to operational stability. Future research should focus on molecular design innovations, such as incorporating reversible conformational switches or thermally responsive functional units. Only through such strategies can the limitations of single-field modulation be overcome, fully harnessing the potential of thermal fields in the practical implementation of molecular rectifiers.

#### *3.4.3 Optical Field Modulation*

Light offers an attractive pathway for regulating molecular rectifiers.[154] Compared with conventional external stimuli such as electric fields or temperature, optical control provides rapid response, high selectivity and remarkable reversibility.[155] In typical designs, photoresponsive molecules such as azobenzene, diarylethene and spiropyran are incorporated as the active units, enabling programmable transport under irradiation with specific wavelengths (**Fig. 8a)**.[156] The underlying mechanisms include photoisomerization, photoinduced charge transfer and photoinduced charge accumulation, among which photoisomerization has emerged as the most widely exploited and effective strategy.[157]

Photochromic molecules often possess two or more stable conformations that can undergo reversible interconversion under ultraviolet or visible light.[158] Such structural switching alters the molecular energy alignment and its coupling with electrodes, leading to pronounced changes in conductance and RRs. The underlying mechanisms include photoisomerization, photoinduced electron transfer, and light-driven charge accumulation. A representative example is the symmetric single-molecule Ru-diarylethene (Ru-DAE) junction reported by Xin and colleagues (**Fig. 8b**).[28] Under UV/visible irradiation, Ru-DAE switches between closed- and open-ring configurations, and in the presence of a gate field its RRs reached more than two orders of magnitude. This approach breaks away from the conventional reliance on chemical asymmetry, opening a new route for device design. Inspired by this concept, Wu and co-workers used first-principles calculations to embed photoresponsive units into molecular rectifiers, revealing optically switchable rectification and elucidating the orbital-level mechanisms.[159] In parallel, Ai and colleagues employed spiropyran as a photoacid to reversibly switch molecular junctions between rectifying and non-rectifying states (**Fig. 8c**).[160] Collectively, these studies extend the scope of optical regulation in molecular electronics and introduce functionalities such as reconfigurable circuits, which are difficult to achieve in conventional semiconductor platforms.

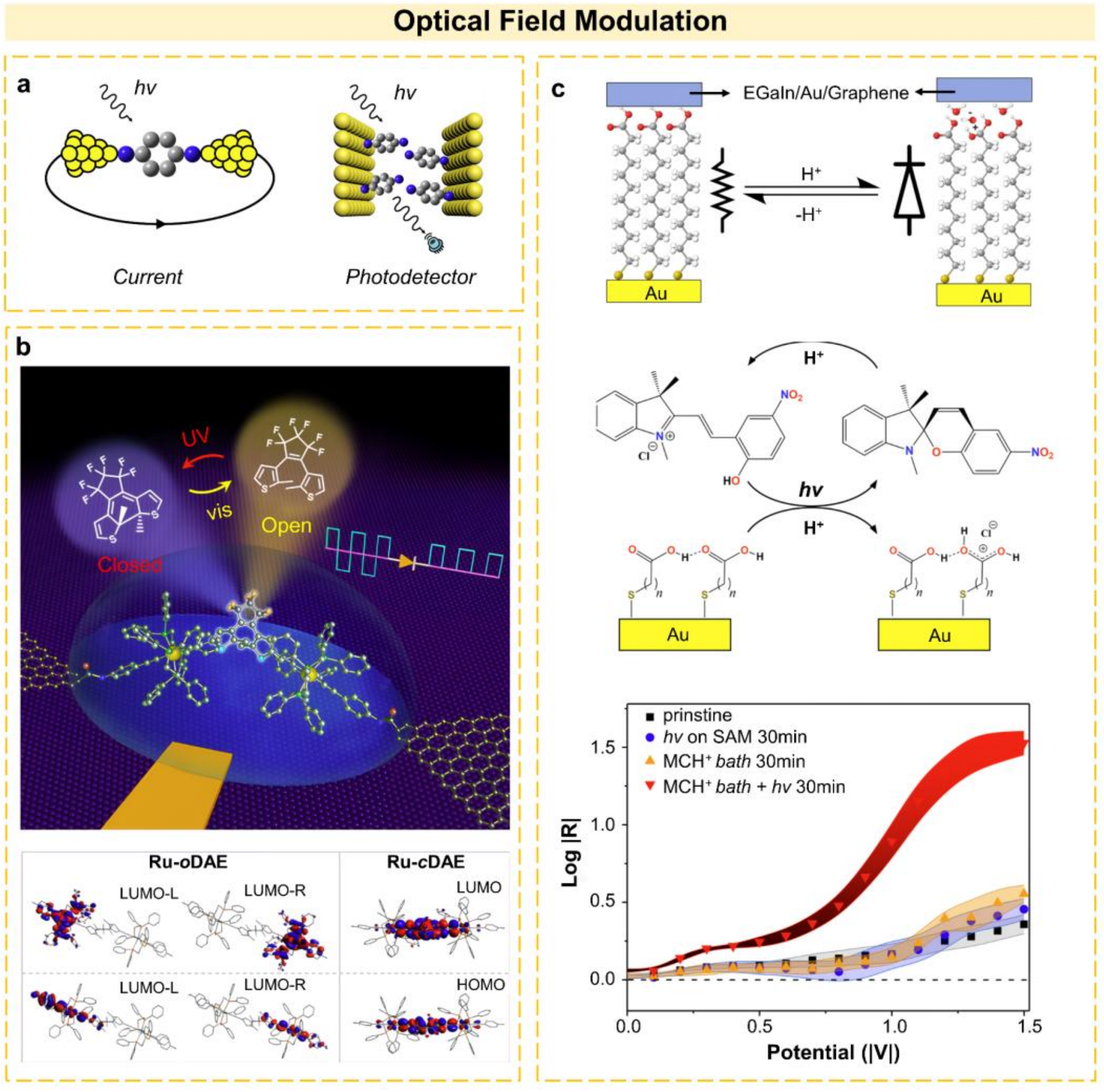

**Figure 8. Optical Field modulation strategies. (a)** Graphical illustration of photo switching molecular junctions simultaneously probed by tunneling current (left) and optical signals (right).Reproduced with permission Ref. [156]. Copyright 2024, Elsevier Ltd. **(b)** Graphene-Ru-DAE-graphene single-molecule transistor with an ionic liquid gate. The schematic highlights the light-stimulated diarylethene (DAE) isomerization between closed-ring (Ru-*c*DAE) and open-ring (Ru-*o*DAE) forms, accompanied by distinct shifts in their FMO diagrams. Reproduced with permission Ref. [28]. Copyright 2021, American Chemical Society. **(c)** Photo-controlled rectification in $Au^{TS}/S(CH_2)_{11}CO_2H$//EGaIn junctions. Light triggers the conversion of protonated merocyanine ($MCH^+Cl^-$) to spiropyran (SP) releases HCl, protonating the $CO_2H$-terminated SAM. The plots of log $|R|$ versus $|V|$ for demonstrate the resulting transition from non-rectifying to rectifying states. Reproduced with permission Ref. [160]. Copyright 2018, American Chemical Society.

A further unique advantage of optical control lies in its inherent visibility. Photoinduced structural changes often generate characteristic absorption bands in the visible spectrum, accompanied by perceptible color variations that greatly facilitate experimental readout.[161, 162] Yet significant challenges remain. Intense irradiation may cause irreversible molecular degradation, whereas insufficient light intensity may fail to fully drive isomerization. Balancing stability with efficiency, and developing photoactive scaffolds with enhanced fatigue resistance and higher switching efficacy, will be essential for advancing optically controlled molecular rectifiers.

#### *3.4.4 Chemical Environment Modulation*

Chemical environments can regulate or enhance molecular rectification by altering the surrounding medium, such as polar solvents, ionic solutions, or humidity.[163] The key mechanism is the reshaping of the local electrostatic landscape and energy-level alignment, which induces asymmetry in charge transport. Unlike chemical gating that relies on external voltages and complex multi-electrode architectures, this strategy exploits dynamic environmental variations, offering a simple but powerful route to control rectification. Such features are particularly attractive for miniaturized and biocompatible molecular devices.[164]

Polar media drive rectification by reconstructing the electrode-molecule interface, forming asymmetric double layers or selective ion accumulation. A representative study by Capozzi et al. exploited the large disparity in exposed surface area between probe (~1 $\mu m^2$) and substrate (>1 $cm^2$) electrodes to generate an asymmetric double layer in propylene carbonate (PC) solution (**Fig. 9a**).[165] Under negative bias, the LUMO approached the probe Fermi level, more orbitals entered the bias window, and current increased sharply. Under positive bias, the LUMO shifted away, limiting orbital participation and suppressing current. This configuration delivered RRs above 200 at voltages as low as 370 mV. Control experiments showed that rectification emerged only in polar solvents (PC, ionic liquids, water) and vanished in non-polar ones such as 1,2,4-trichlorobenzene (TCB). Even symmetric molecules like 4,4′-bipyridine exhibited strong rectification in polar environments. These results highlight the universality and efficiency of ion-induced interfacial reconstruction, further supported by theoretical analysis.[166]

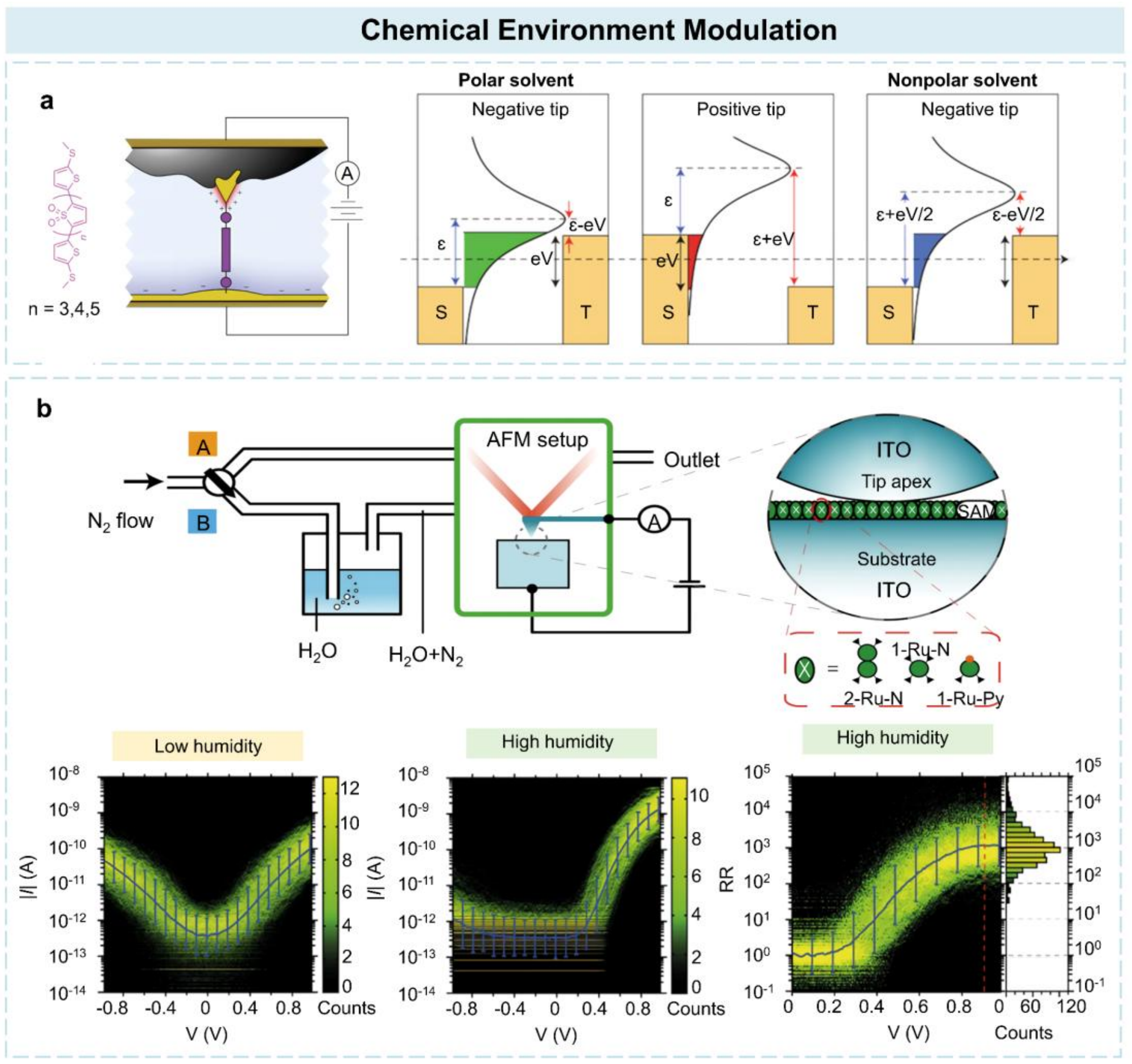

**Figure 9. Chemical Environment Modulation Strategies.** (a) Rectification mechanism in TDOn molecular junctions. The schematic and energy-level diagrams illustrate the influence of solvent polarity on orbital alignment. In polar media, the molecular resonance remains pinned to the substrate chemical potential, leading to asymmetric bias window overlap and high rectification. In non-polar solvents, the resonance position is bias-independent, resulting in symmetric transport. Reproduced with permission Ref. [165]. Copyright 2015, Springer Nature. (b) Humidity-modulated charge transport in ITO-molecule-ITO junctions. The conductive-probe atomic force microscopy (C-AFM) setup allows for precise control of the $N_2$ environment. The 2D histograms of logarithmically binned *I-V* characteristics and RR for 2-Ru-N molecular junctions demonstrate the transition from low-conductance states at 5% humidity to high-performance rectifying states at 60% humidity. Reproduced with permission Ref. [167]. Copyright 2018, Springer Nature.

Humidity offers another effective variable. In monolayers of di-nuclear ruthenium complexes, raising relative humidity from 5% to 60% enabled reversible switching between symmetric and asymmetric I-V responses (**Fig. 9b**).[167] The underlying mechanism is humidity-dependent orbital misalignment. In dry conditions, two serially coupled orbitals are nearly degenerate, whereas high humidity induces displacement of $PF_6^-$ counterions, breaking the degeneracy and triggering orbital mismatch, thereby reshaping the tunneling barrier. This demonstrates that humidity control shares the same physical origin as ionic solution regulation, both rooted in environment-mediated interfacial charge reconstruction.

At a more refined level, Debye screening in electrolyte solutions provides a quantitative handle for rectification tuning. As early as 2006, Kornyshev and co-workers theoretically predicted that RRs increase with molecular length and electrolyte concentration.[168] In dilute electrolytes, where the Debye length far exceeds the molecular length, rectification vanishes and current-voltage (I-V) responses become symmetric. Recent experiments on iodo-terminated hexathiophene molecule (IT-6) and circumanthracene-based dibenzonitrile molecule (CABD) molecular junctions confirmed this prediction.[27] In NaF aqueous solutions, increasing ionic concentration shortened the Debye length and systematically enhanced rectification.

Overall, chemical environment regulation offers a general and elegant route to improve molecular rectification. However, experimental outcomes are highly sensitive to solvent properties and humidity, while existing theoretical frameworks are insufficient to fully rationalize the diverse responses observed across molecular systems. Future efforts should focus on establishing quantitative relationships between interfacial structure and environmental factors, paving the way from empirical adjustments toward predictive and robust design of molecular functionalities.

# 4 Fabrication Techniques

The fabrication of stable molecular junctions remains the central challenge in the field of molecular rectifiers.[169] Critical hurdles include the reliable creation of nanogap electrodes, the precise incorporation of molecules into these gaps, and the realization of rectification devices with both high performance and reproducibility. Current strategies for constructing molecular rectifiers can be broadly classified into two categories, distinguished by whether the electrodes remain fixed or undergo repeated motion. The first involves dynamic junction techniques, where controlled opening and closing of electrodes allows the repeated formation of molecular contacts.[51, 170] The second category relies on static junction techniques, in which fixed electrodes are used to define molecular junctions.[171]

## 4.1 Dynamic Break Junction Methods

Dynamic break junction techniques enable the reversible formation and rupture of molecular junctions by continuously tuning the electrode spacing. This approach allows the construction of single-molecule rectifiers in the strict sense, while repeated opening and closing of the metallic electrodes permits thousands of junction formation events. Among these methods, MCBJ and STM-BJ have emerged as the dominant platforms for fabricating dynamic single-molecule junctions. [51, 170, 172, 173] Other variations, such as atomic force microscopy break junctions (AFM-BJ) and microelectromechanical system break junctions (MEMS-BJ), have also been reported and can be found in specialized reviews.[174, 175]

### *4.1.1 Mechanically Controllable Break Junction*

Since its introduction in 1997, the MCBJ technique has been one of the most representative and powerful tools in molecular electronics (**Fig. 10a**).[51] Its principle is conceptually simple but technically precise: a metallic nanowire is deposited on a flexible substrate, and a pushing rod bends the substrate until the nanowire elongates and eventually fractures. This process generates atomically sharp electrodes separated by a controllable sub-nanometer gap. Molecules equipped with anchoring groups can then bridge the gap, forming stable single-molecule junctions whose transport properties can be directly probed.

A landmark study appeared in 2005, when Elbing et al. employed the MCBJ technique to demonstrate molecular rectification at the single-molecule level.[82] They incorporated donor-acceptor substituted molecules into gold nanogaps and observed diode-like I-V characteristics with RRs approaching 10 at ±1.5 V. This work provided the first experimental confirmation of the AR model and highlighted the decisive role of molecular orbital alignment in rectification. From that point on, MCBJ became a central experimental platform for probing single-molecule diodes, laying the foundation for uncovering the intrinsic mechanisms of rectification. Beyond establishing feasibility, MCBJ has enabled systematic investigations into structure-function relationships in molecular rectifiers. Researchers have exploited its stability and precision to unravel how molecular backbone length, substituent type, intermolecular stacking, and conformational dynamics dictate rectification behavior.[49, 176] More recently, MCBJ has been coupled with external field modulation, particularly electrostatic gating, to dynamically shift molecular orbital energies.[143] Such experiments opened the way to exploring gate-tunable rectification and even introduced the conceptual framework of “molecular field-effect transistors”, thereby deepening our understanding of charge transport under multi-field coupling (**Fig. 10b**).

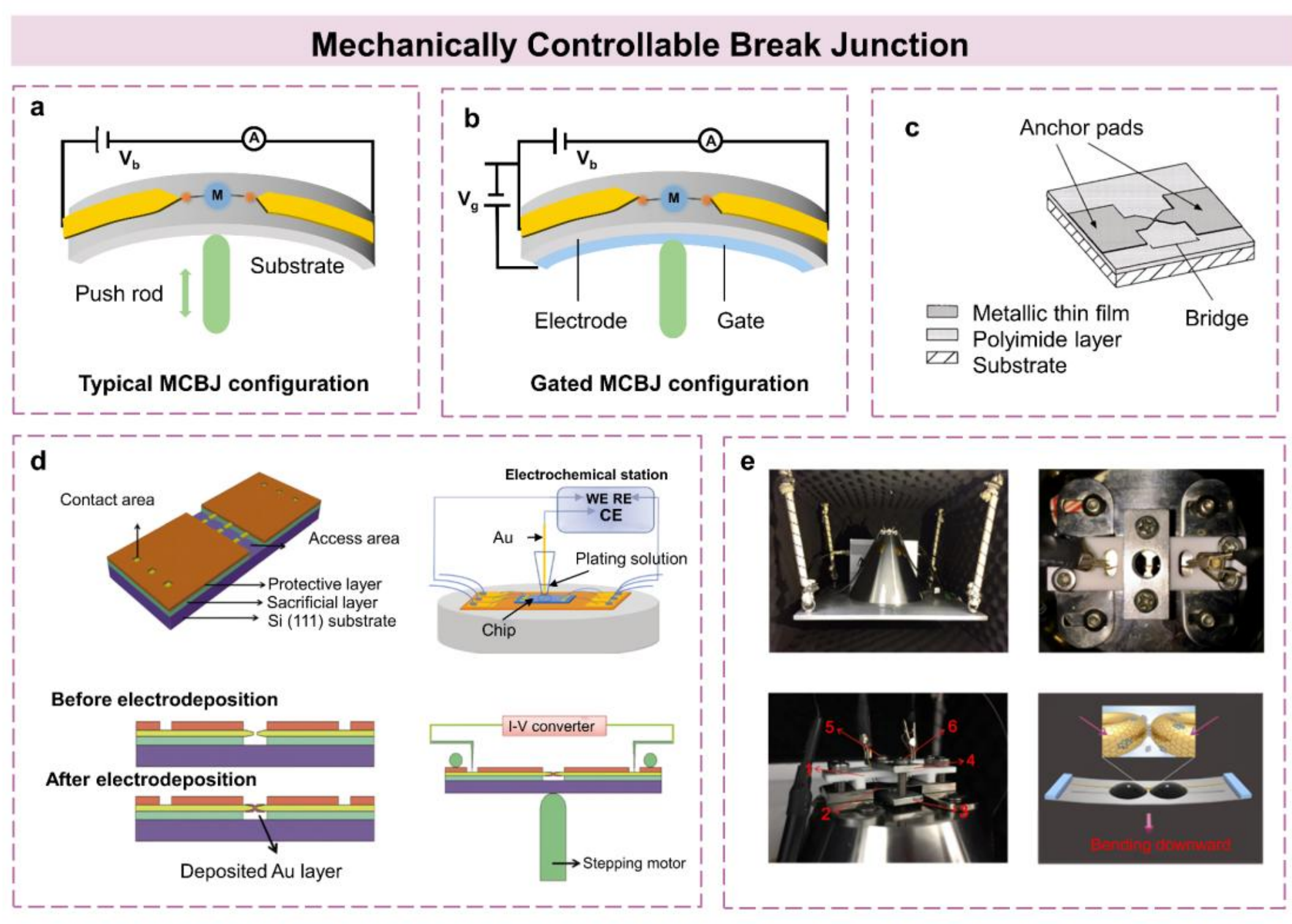

**Figure 10. Mechanically Controllable Break Junction.** **(a)** Typical MCBJ configuration. Bending a flexible substrate yields atomically sharp electrodes with controllable gaps. **(b)** Gated MCBJ device. **(c)** Schematic of chip fabrication. A polyimide layer acts as both an insulating and sacrificial layer, with a central "bridge" neck acting as a strain concentrator for electrode formation. Reproduced with permission Ref. [178]. Copyright 1996, American Institute of Physics. **(d)** Electrochemically assisted MCBJ (EC-MCBJ). The chip on a Si (111) substrate undergoes precise gold electrodeposition within an electrochemical station (WE, RE, and CE) to create nanobridges. The system integrated with an I-V converter and stepping motor enables high-throughput single-molecule conductance measurements. Reproduced with permission Ref. [177]. Copyright 2011, Springer Nature. **(e)** Cross-plane break junction (XPBJ) experimental setup. The photos and schematics illustrate the device assembly comprising a liquid cell (1), a graphene chip (2), two alloy plates (3, 4), and two screws acting as adapting pieces (5, 6), where the downward bending of the graphene chip enables the investigation of single-molecule conductance. Reproduced from Ref. [179] under the CC BY 4.0 license.

Advances in nanofabrication have also extended MCBJ beyond traditional metallic substrates. Variants such as electrochemically assisted MCBJ chips,[177] nano-lithographically defined junctions,[178] and carbon-based MCBJ platforms have been reported, providing improved stability for long-term measurements (**Fig. 10c and d)**. A striking example is the cross-plane break junction (XPBJ) method pioneered by Hong and co-workers, as shown in **Fig. 10e**, which uses CVD-grown graphene transferred onto metallic wires.[179] In this configuration, van der Waals interactions rather than conventional anchoring groups stabilize the molecular junction. Remarkably, RRs as high as 457 were achieved in polycyclic aromatic hydrocarbons without chemical linkers, demonstrating the potential of all-carbon electrodes for molecular rectification.[113]

Taken together, MCBJ has played a foundational role in the evolution of single-molecule rectifiers. It not only enabled the first experimental realization of rectification at the molecular scale but continues to drive the exploration of both structural tuning and external-field modulation of transport. Although limitations remain notably the inability to image electrode-molecule interfaces in real time, its exceptional stability, controllability, and ability to probe intrinsic mechanisms ensure that MCBJ remains an indispensable platform for advancing molecular rectifier research.

#### *4.1.2 Scanning Tunneling Microscope Break Junction*

The STM-BJ technique originates from scanning tunneling microscopy (STM). With a piezoelectric actuator capable of controlling the probe-substrate distance at the ångström (Å) scale, STM-BJ offers unparalleled precision in single-molecule imaging and manipulation. Tao et al. first introduced this approach to single-molecule conductance measurements, successfully probing the conductance of bipyridine (BPY) junctions through repeated stretching-releasing cycles.[170] This breakthrough not only opened a new avenue for single-molecule charge transport studies but also laid the experimental foundation for constructing molecular rectifiers.[83]

Unlike the lateral breaking process of MCBJ, STM-BJ relies on precise vertical motion of the STM tip to form molecular junctions. During operation, the tip may directly contact the substrate or remain in weak interaction or even non-contact states (**Fig. 11a**, hard-contact and soft-contact mode). Through rapid and repeatable stretching-releasing cycles, thousands of junctions can be constructed and statistically analyzed within a short time. Such high-throughput capability is particularly powerful, making STM-BJ a key platform for uncovering universal rectification behaviors and probing underlying transport mechanisms.

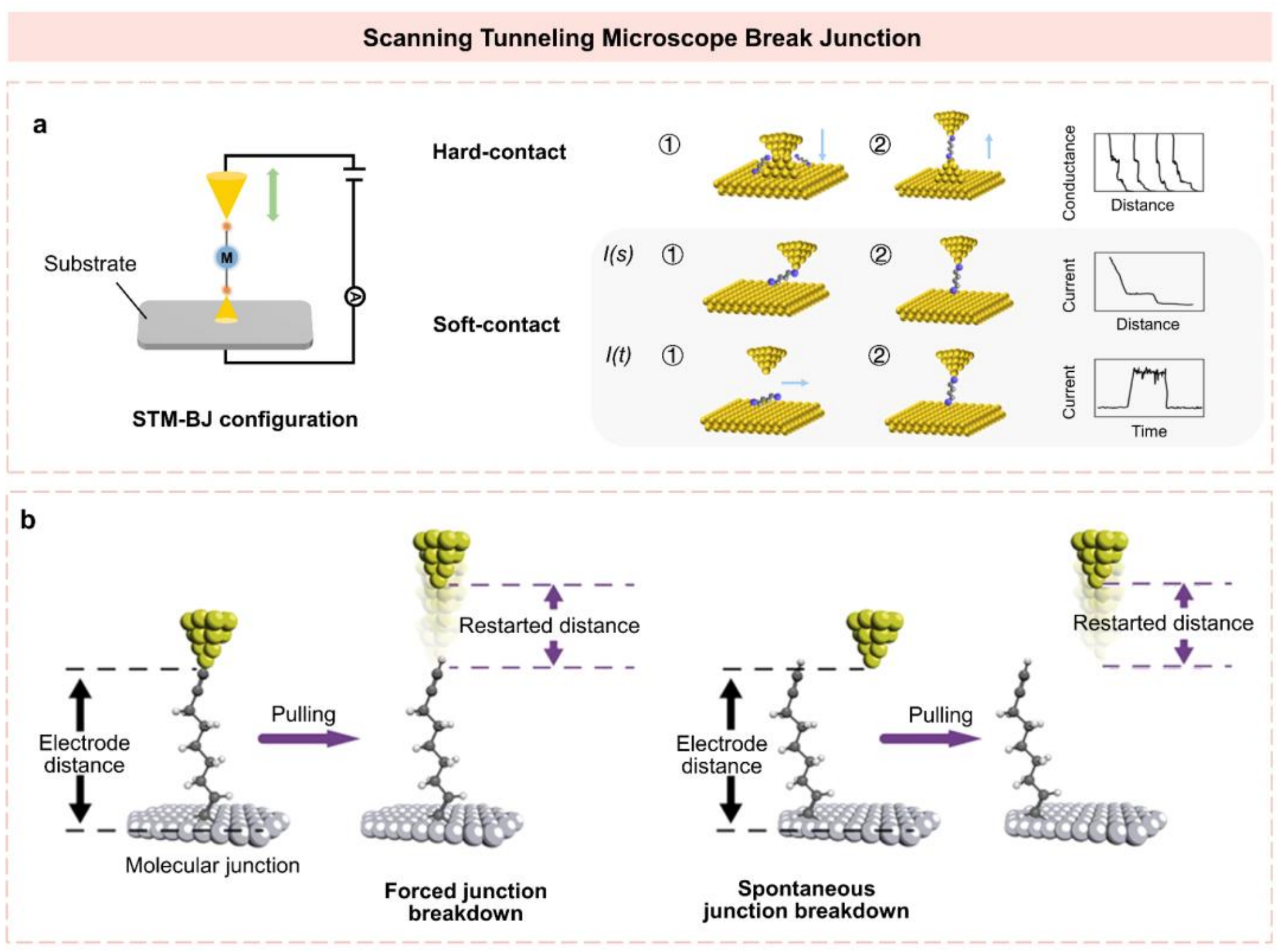

**Figure 11. Scanning tunneling microscope break junction (STM-BJ) techniques. (a)** Schematic of the STM-BJ setup and molecular junction evolution modes. The illustrations compare the formation and breaking processes under hard-contact mode, *I*(*s*) mode, and *I*(*t*) mode. Reproduced from Ref. [175] under the CC BY 4.0 license. **(b)** STM break-junction experiments on asymmetric gold-nonadiyne-silicon junctions. The schematics depict the pulling stage of a blinking event, contrasting the junction configurations before and after spontaneous junction breakdown occurs. Reproduced from Ref. [183] under the CC BY 4.0 license.

In molecular rectifier studies, STM-BJ is most commonly employed to fabricate metal-molecule-metal junctions.[102, 180] Owing to the introduction of the probe, the electrode geometry and electronic structure often exhibit inherent asymmetry, which in certain cases amplifies the rectification effect. To broaden its applicability, researchers have progressively expanded the range of electrode materials. Beyond the conventional choices of gold (Au), silver (Ag), and platinum (Pt), electrochemical deposition has enabled the integration of other metals such as copper (Cu) and iron (Fe) onto the STM tip, thereby enriching the electrode palette.[181, 182] Meanwhile, semiconductors including silicon (Si) and gallium arsenide (GaAs) have been incorporated as substrate electrodes to construct asymmetric molecular junctions.[183, 184] This strategy not only diversifies electrode selection but, more importantly, introduces band-structure modulation as an additional lever for tuning rectification performance. For example, in single-nonadiyne molecular junctions, Au top electrodes were combined with N-doped Si bottom electrodes, where the molecule formed highly controlled covalent bonds and spatial configurations on the Si surface (**Fig. 11b**).[183] The resulting devices exhibited superior reproducibility and mechanical stability compared with conventional Au-based junctions. Remarkably, with low-doped Si electrodes, RRs exceeded 4,000, underscoring the potential compatibility of single-molecule junctions with semiconductor technologies. Similarly, carbon-based electrodes have been integrated into STM-BJ platforms. For instance, Kim et al. employed graphite as the bottom electrode to construct molecular junctions, opening a new avenue for developing heterogeneous molecular rectifiers using layered two-dimensional materials.[112]

Compared with MCBJ, STM-BJ does not require sophisticated chip fabrication and involves a relatively simple experimental setup, making it particularly suitable for large-scale statistical studies with high reproducibility.[106, 185-187] However, unavoidable mechanical interactions between the tip and molecules remain a challenge, limiting the mechanical stability and precise control of the electrode-molecule interface in rectifier applications. Despite these limitations, the high-throughput capability, flexibility, and material compatibility of STM-BJ have firmly established it as a key experimental methodology driving mechanistic insights and performance optimization in molecular rectifiers.

### 4.2 Static Molecular Junction Techniques

Static molecular junctions are constructed by assembling target molecules within a fixed electrode gap. Compared with dynamic junctions, this strategy eliminates the need for continuous control of electrode spacing, offering higher structural stability and better compatibility with semiconductor processing. Static molecular junctions are generally classified into three types: self-assembled monolayers (SAMs), carbon-based junctions, and electromigrated break junctions (EBMJs).[28, 188, 189] Among them, SAMs and carbon-based junctions show clear advantages in reproducibility, interfacial control, and device scalability, and will be the focus of this review. Detailed progress on EBMJs has been summarized elsewhere.[190]

#### *4.2.1 Self-assembled Monolayers*

SAM-based junctions represent a highly controllable and stable strategy for ensemble molecular devices. The principle relies on anchoring groups that self-assemble into an ordered monolayer on the bottom electrode, followed by deposition of a top electrode to complete the junction, as depicted in **Fig. 12a**. Although this method does not provide precise positioning at the single-molecule scale, it offers significant benefits in device uniformity, thermochemical stability, and large-scale integration. These features have made SAMs a central platform for reproducible and scalable studies of molecular rectifiers.

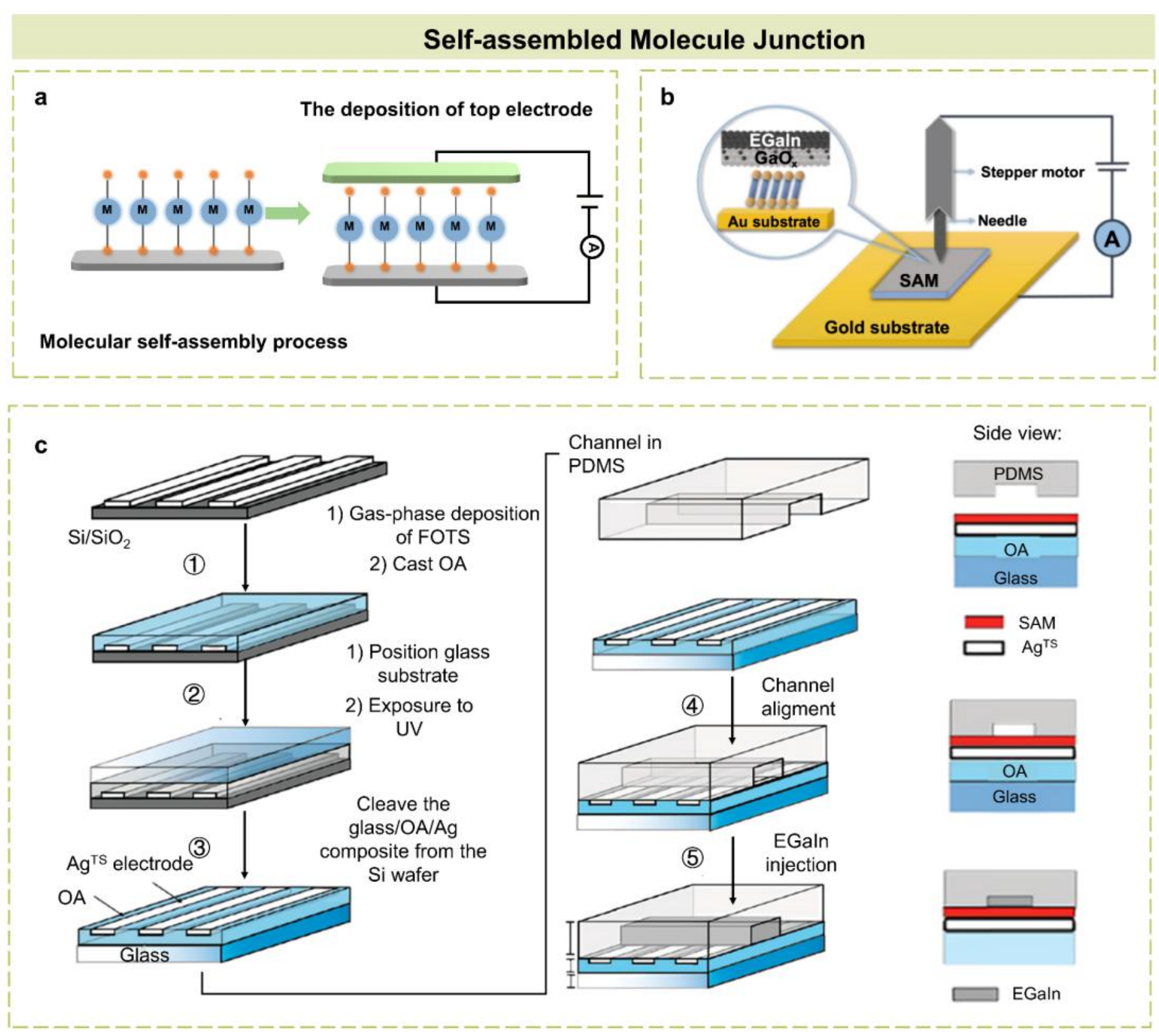

**Figure 12. SAM-assembled molecule junctions. (a)** Typical SAM-based molecular junction. The architecture relies on anchoring groups that self-assemble into an ordered monolayer on a bottom electrode, followed by the soft-contact deposition of a top electrode to complete the junction. **(b)** Eutectic gallium-indium (EGaIn) /SAM/Au junction. The liquid metal electrode conforms to the SAM surface, providing a non-destructive contact that avoids short-circuit issues. Reproduced with permission Ref. [195]. Copyright 2019, Wiley-VCH. **(c)** Fabrication of the arrays of SAM-based tunneling junctions. The sequential process illustrates the template-stripping of silver electrodes to achieve ultra-flat surfaces, followed by the alignment of a microchannel fabricated from a transparent polymer (PDMS) and subsequent filling with $Ga_2O_3$/EGaIn to form high-throughput junction arrays for statistical characterization. STM break-junction experiments on asymmetric gold-nonadiyne-silicon junctions. Reproduced with permission Ref. [196]. Copyright 2010, American Chemical Society.

Early attempts to fabricate SAM junctions relied on metal deposition such as Au or Ti.[191] This often punctured the organic monolayer, causing short circuits and poor reproducibility.[192] To overcome this challenge, researchers explored alternative strategies, including the introduction of buffer layers such as graphene, conductive polymers, or ultrathin $Al_2O_3$, as well as soft-contact methods such as transfer printing or nanoimprinting.[193, 194] Yet these approaches generally suffered from uncontrollable interfaces, low yield, and high experimental complexity, and therefore failed to gain widespread adoption. A true breakthrough came with the introduction of eutectic gallium-indium (EGaIn) electrodes. Owing to its low surface tension and excellent fluidity, EGaIn can smoothly conform to SAMs without damaging the molecular layer, effectively avoiding the puncturing problem inherent to metal deposition. A typical process involves assembling a functional SAM on a gold substrate, followed by gentle contact with liquid EGaIn as the top electrode (**Fig. 12b**).[195] Building upon this soft-contact advantage, researchers have developed sophisticated fabrication protocols that integrate template-stripping techniques with microfluidic systems to enable the scalable production of high-throughput junction arrays (**Fig. 12c**).[196] This automated assembly process not only minimizes manual handling errors but also allows for the simultaneous characterization of hundreds of molecular junctions, providing a robust statistical foundation for analyzing molecular electronic properties. Since its introduction, EGaIn has transformed SAM-based junctions into a reproducible and scalable experimental platform. Reported RRs typically fall in the range of 10 to $10^3$, and can even exceed $10^5$ with careful molecular design and interfacial optimization, rivaling or surpassing certain semiconductor devices.[114, 197, 198] Importantly, the EGaIn platform enables statistical validation and quantification of molecular rectification, greatly enhancing the reliability and comparability of experiments.

Nevertheless, SAM junctions are not without limitations. Variations in molecular orientation, packing density, and defects inevitably introduce device-to-device variability. Environmental sensitivity, such as humidity and oxidation, can further compromise stability. More critically, the technique is intrinsically limited in spatial resolution and reflects only the averaged transport behavior of molecular layers, rather than the intrinsic properties of individual molecules. As such, SAM-based junctions are better suited for ensemble studies of molecular rectification than for probing single-molecule transport. To further transcend the limitations of molecular orientation variability and grain boundary defects inherent in traditional SAMs, monolayer organic crystals (MOCs) have recently emerged as a highly promising platform for static molecular junctions. Unlike the stochastic assembly of SAMs, MOCs provide a long-range ordered lattice that ensures exceptional uniformity and intrinsic stability. For instance, Li et al. demonstrated the growth of high-quality monolayer $C_6$-DNTT single crystals over millimeter scales using solution-shearing epitaxy.[26] By integrating these MOCs with van der Waals (vdW) metal contacts, they fabricated molecular diodes that achieved a record-high rectification ratio of $5 \times 10^8$ and an ideality factor close to unity. This achievement not only sets a new benchmark for rectification ratios in molecular junctions but also provides a promising strategy to mitigate the long-standing issues of reproducibility and stability that have plagued single-molecule devices, thereby complementing the limitations of SAM-based platforms.

#### *4.2.2 Carbon-based Junctions*

Carbon-based molecular junctions have emerged as a promising alternative that overcomes key limitations of metal electrodes. Carbon nanomaterials such as carbon nanotubes (CNTs) and graphene combine unique electronic structures with high interfacial compatibility and mechanical flexibility.[199-201] These features reduce interfacial defects and suppress MIGS, allowing molecular level alignment that is closer to theoretical predictions. In contrast to SAMs, carbon-based junctions can more readily achieve static single-molecule configurations, providing a new experimental route to explore intrinsic charge transport.

Fabrication of carbon-based junctions typically involves creating nanogaps within carbon electrodes, followed by the insertion of functional molecules. Common approaches include electrical breakdown, local oxygen plasma etching, and focused ion beam (FIB) milling, with oxygen plasma etching widely used due to its simplicity and controllability.[202-204] Among these, the physical strategy of current-induced breakdown allows for the creation of nanogaps that can be further refined through electron-beam irradiation, as demonstrated in the fabrication of DNA-based molecular devices (**Fig. 13a**). This method facilitates in situ observation and precise gap adjustment to match the dimensions of the target biomolecules. Plasma treatment generates nanogaps terminated with carboxyl groups, which can covalently link to amine-terminated molecules via amide bonds, ensuring stable molecular integration. More recently, the combination of plasma etching with lithography and electrical breakdown has enabled high-yield fabrication of CNT and graphene nanogap electrode arrays, laying the foundation for large-scale integration of molecular rectifiers.[205, 206]

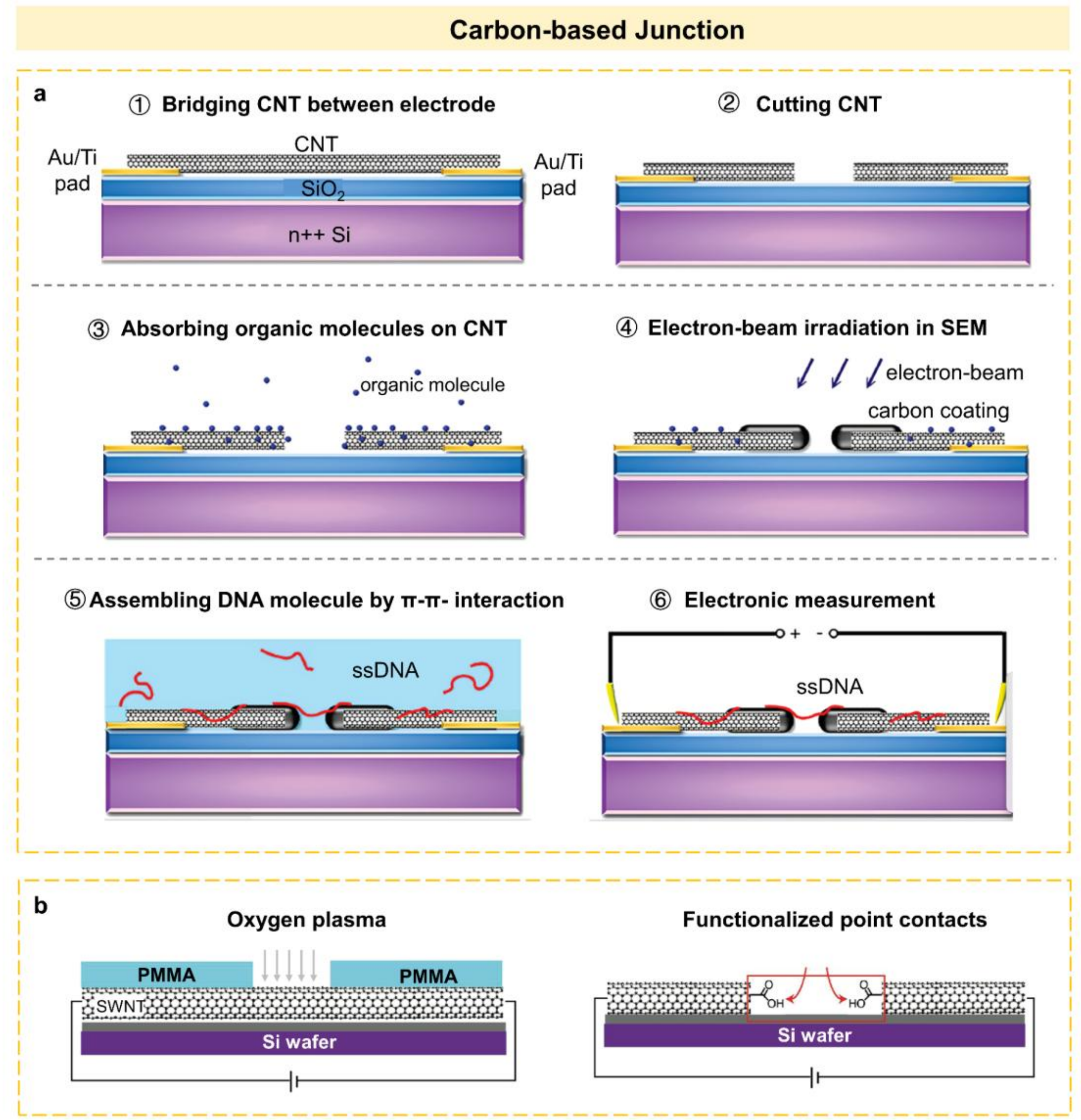

**Figure 13. Carbon-based junctions. (a)** Fabrication of carbon nanotube (CNT) electrodes with controlled nanogaps for DNA electronic devices: the process involves bridging a CNT between Au/Ti electrodes, followed by cutting via current breakdown and electron-beam irradiation to refine the gap, enabling the assembly and electronic measurement of DNA molecules. Reproduced with permission Ref. [202]. Copyright 2008, American Chemical Society. **(b)** Precision cutting of single-walled carbon nanotubes (SWNTs) via oxygen plasma. A sub-10 nm window is defined through a PMMA mask using electron-beam lithography, where oxygen plasma etching creates carboxylate-functionalized point contacts for molecular bridging. Reproduced with permission Ref. [207]. Copyright 2008, American Chemical Society.

Taking CNT-based molecular rectifiers as a representative example, a typical chemical fabrication process based on oxygen plasma etching is illustrated in **Fig. 13b**.[207] Large-area CNT is first synthesized by chemical vapor deposition and transferred onto a silicon substrate. A thin layer of polymethyl methacrylate (PMMA) is then spin-coated and patterned using electron-beam lithography to define micro- and nanoscale electrode features. Oxygen plasma etching combined with controlled electrical breakdown generates nanogap point electrodes terminated with carboxyl groups. Target molecules are subsequently assembled into the gap and covalently coupled to the CNT electrodes, yielding a robust molecular rectifier. This strategy provides distinctive advantages in interfacial coupling and energy-level alignment and establishes an experimental paradigm for reproducible and scalable high-performance molecular rectification.

The potential of carbon-based junctions lies in their interfacial compatibility, efficiency of molecular integration, mechanical flexibility, and scalability.[208] However, challenges remain, particularly the complexity of material synthesis and transfer, as well as limited uniformity of large-area electrodes. Despite these hurdles, carbon electrodes mark a paradigm shift in molecular rectifier fabrication, expanding the material landscape and offering a promising pathway toward scalable, high-performance molecular electronics.

# 5 Characterization Techniques

The performance of molecular rectifiers depends not only on the precise synthesis of individual molecules but also on the assembly of the complete device. Consequently, their characterization requires a multiscale strategy spanning from the molecular level to the device scale.[209] In this section, techniques for probing molecular and interfacial structures are first reviewed, followed by methods for quantifying rectification performance via electrical measurements. Finally, approaches for characterizing the electronic structure are discussed. Collectively, these methodologies establish a comprehensive framework for understanding the functional behavior and charge transport processes in molecular rectifiers.

## 5.1 Structure and Interface Characterization

The performance of molecular rectifiers is determined not only by the intrinsic structure of the molecules but also by their interactions with electrode interfaces.[210] Molecular composition and conformation, electrode geometry, the size of nanogaps, and the nature of anchoring and chemical bonding at the interface all directly govern charge transport efficiency and RRs. Consequently, structural and interfacial characterization of molecular rectifiers must span multiple length scales, from the molecular to the device level.

At the molecular level, structural integrity and chemical purity are prerequisites for reliable device fabrication. Commonly employed techniques include nuclear magnetic resonance (NMR) spectroscopy, high-resolution mass spectrometry (HRMS), infrared spectroscopy (IR), and differential scanning calorimetry (DSC).[211-214] NMR provides detailed information on the molecular backbone, functional groups, and the chemical environment of substituents, whereas HRMS validates molecular mass and purity. IR spectroscopy and DSC focus on conformation and thermal stability. IR detects conformational changes induced by external stimuli, such as temperature or light, via characteristic vibrational peaks of functional groups. DSC evaluates the suitability of molecules for device integration by measuring melting ($T_m$) and decomposition temperatures ($T_d$). These methods not only confirm the synthetic fidelity and stability of target molecules but also lay a robust foundation for subsequent interfacial assembly and electrical performance studies.

At the device level, characterization primarily addresses two key aspects: the geometry and dimensional control of nanogap electrodes, and whether molecules form stable and ordered assemblies on the electrode surface. Techniques commonly employed include atomic force microscopy (AFM), scanning electron microscopy (SEM), scanning tunneling microscopy (STM), high-angle annular dark-field scanning transmission electron microscopy (HAADF-STEM), energy-dispersive X-ray spectroscopy (EDS), X-ray photoelectron spectroscopy (XPS), angle-resolved XPS (ARXPS), and ellipsometry.[215-222] AFM, SEM, and STM probe electrode morphology and the spatial distribution of nanogaps at the microscale. HAADF-STEM and EDS reveal elemental composition and structural details of the molecule-electrode interface at the atomic scale. XPS and ARXPS provide insights into interfacial chemical states and the bonding modes of anchoring groups, while ellipsometry precisely measures monolayer thickness and coverage. Together, these complementary techniques map the structural landscape of interfaces across multiple scales.

The focus of characterization is strongly influenced by device fabrication strategies. In SAM systems, molecular order and film uniformity are critical for device stability. AFM is indispensable for assessing monolayer height distribution and local defects, while ellipsometry rapidly evaluates large-area film thickness and coverage. XPS serves as core tool for confirming chemical bonding at the interface. When additional information on molecular stacking orientation or long-range order is required, X-ray diffraction (XRD) can provide valuable insight. [102, 223] Besides, near-edge X-ray absorption fine structure (NEXAFS) spectroscopy can sensitively probe the tilt angles and backbone orientation of molecules, offering complementary atomic-scale information critical for understanding the conformation at molecule-electrode interfaces.[224] As exemplified in **Fig. 14a**, it presents the molecular structures of n-mercaptoalkanoic acids and the device architecture of a molecular junction composed of a C11 SAM on an Au substrate with a $Ga_2O_3$/EGaIn top electrode.[225] **Fig. 14b** shows AFM topography images comparing the bare Au substrate and the 11-mercaptoalkanoic acid (C11) SAM, clearly illustrating the formation of a uniform molecular layer. Ellipsometry data (**Figs. 14c** and **f**) exhibit a linear dependence of film thickness on chain length and a slight increase in thickness after deprotonation. Reflection-Absorption Infrared Spectroscopy (RAIRS) in the C–H and C=O regions (**Figs. 14d** and **g**) confirms the crystalline-like molecular ordering and the conversion of COOH to $COO^-$ groups. Additionally, high-resolution C 1s and K 2p XPS spectra (**Figs. 14e** and **h**) reveal the chemical composition of the terminal groups and the stoichiometric coordination of alkali metal ions after deprotonation. Thus, the combination of AFM, ellipsometry, and XPS constitutes the most common characterization "trio" for SAM-based molecular rectifiers.

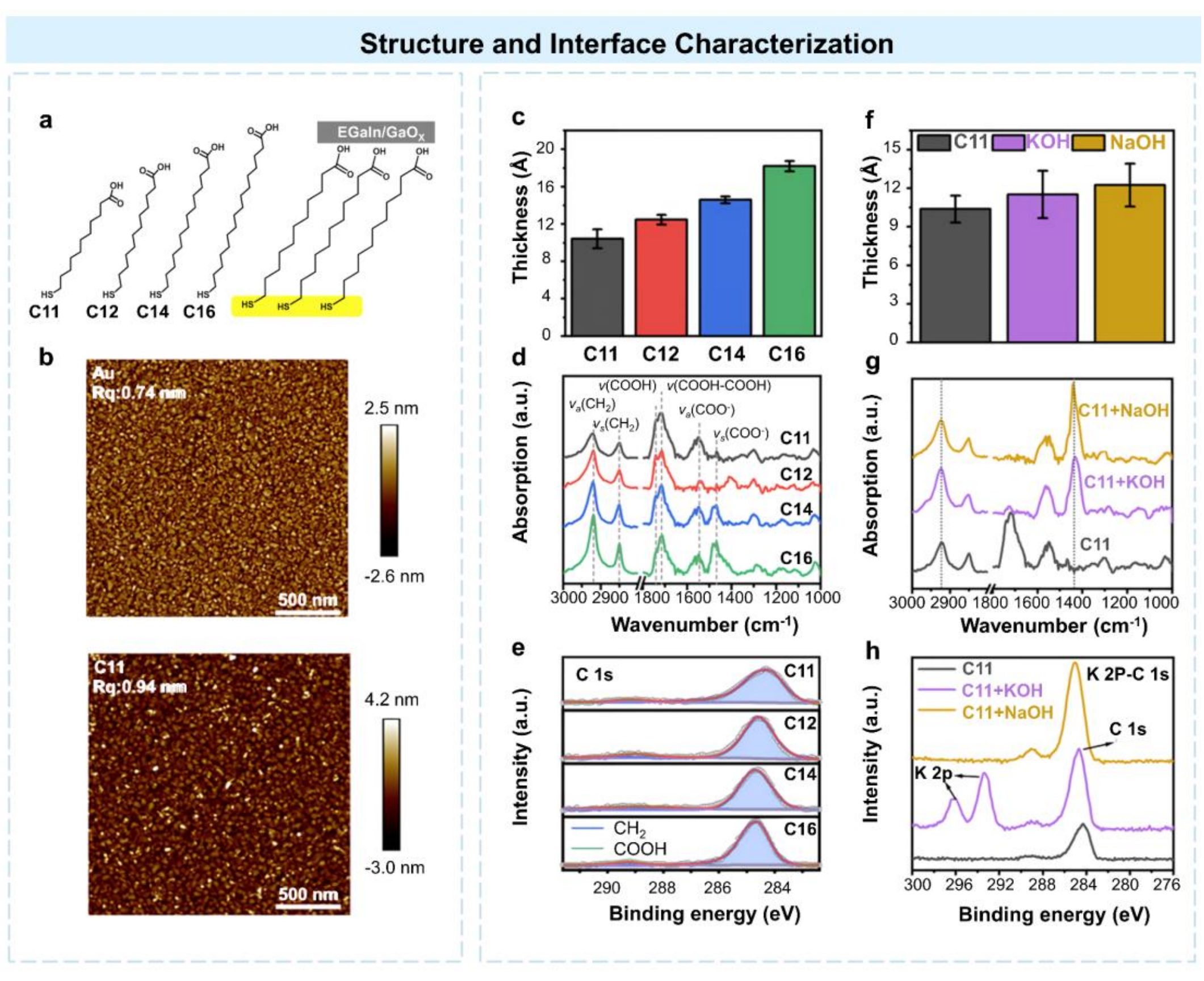

**Figure 14. Structure and interface characterization. (a)** Molecular structures and device architecture. The series includes n-mercaptoalkanoic acids and a schematic illustration of the molecular junction formed by a C11 SAM on an Au substrate with a $Ga_2O_3$/EGaIn top electrode. **(b)** AFM topography images compare the bare Au substrate and the 11-mercaptoalkanoic acid (C11) SAM, illustrating the formation of a uniform molecular layer. **(c, f)** Ellipsometry data showing the linear dependence of film thickness on chain length and the slight thickness increase after deprotonation (**d, g**) Reflection-Absorption Infrared Spectroscopy (RAIRS) in the C–H and C=O regions confirming the crystalline-like molecular ordering and the conversion of COOH to $COO^-$ groups. **(e, h)** high-resolution C 1s and K 2p XPS spectra illustrating the chemical composition of the terminal groups and the stoichiometric coordination of alkali metal ions after deprotonation. Reproduced with permission Ref. [225]. Copyright 2021, American Chemical Society.

For carbon-based electrodes, characterization priorities differ. The structural integrity and defect density of graphene or carbon nanotubes critically influence molecular-electrode coupling. Raman spectroscopy (RS) becomes indispensable for evaluating layer number, defect density, and local strain, whereas SEM and AFM remain essential for verifying surface uniformity and molecular coverage.[48, 56]

In contrast, in break-junction devices prepared via STM-BJ or MCBJ, the interface is not statically stable but undergoes continuous formation-rupture-reformation during electrical measurements. In such systems, conventional imaging and spectroscopic characterization are of limited utility; instead, statistical analysis of thousands of conductance traces is often employed to extract thermodynamically stable molecular configurations. Here, electrical measurements themselves serve as the primary structural probe, with microscopy and spectroscopy acting as auxiliary validation tools.

Overall, structural and interfacial characterization of molecular rectifiers has evolved from single-technique verification to a multi-scale, multi-dimensional analytical framework. Looking forward, the development of in situ dynamic characterization techniques, capable of capturing real-time molecular conformations and interfacial states under operational conditions, is essential.[175, 226, 227] Such advances will enable a deeper understanding of the dynamic mechanisms underlying rectification and provide a robust physicochemical basis for the rational design of molecular electronic devices.

### 5.2 Electrical characterization

The functional performance of molecular rectifiers is ultimately manifested in their electrical response, making electrical characterization a critical step for evaluating device rectification and application potential. Even minor variations in molecular structure or interfacial configuration can induce significant changes in conductance, and molecular junctions constructed via different methods may exhibit markedly distinct rectification behaviors.[228] Establishing standardized and reproducible protocols is therefore essential to enable meaningful comparisons and mechanistic insights.

In dynamic break junction techniques, conductance histograms are typically employed first to identify stable molecular configurations, followed by I-V measurements under those conditions.[170] A stable conductance plateau is a prerequisite for reproducibility and statistical reliability in single-molecule junctions. By repeatedly forming and breaking the junction, thousands of conductance traces can be rapidly collected, generating one-dimensional (1D) or two-dimensional (2D) conductance histograms (**Figs. 15a-c**).[229, 230] A 1D histogram displays the frequency of different conductance channels, with the horizontal axis commonly plotted as $\log(G/G_0)$ (**Fig. 15b**);[231] peak positions correspond to the most stable molecular conformations or electrode-molecule binding motifs, providing a standard reference for typical conductance states. 2D histograms, such as conductance-distance maps, offer additional insight into junction stability during elongation, revealing correlations between molecular orientation, electrode polarity, and rectification behavior (**Fig. 15c**).[230]Once stable conductance plateaus are observed, voltage sweeps are applied to record I-V characteristics (**Fig. 15d**). By aggregating numerous traces into I-V histograms and fitting the data statistically, an average response can be obtained, from which RRs under positive and negative bias are calculated (**Fig. 15e**).

In contrast, static junctions generally involve direct I-V measurement after device fabrication (**Fig. 15f**). Fixed electrode spacing minimizes selection bias associated with stretching trajectories, enabling each measurement to reflect device characteristics. However, as the specific molecular configuration is uncontrolled, measurements may only represent certain junctions. Consequently, repeated measurements across multiple devices, combined with Gaussian fitting or similar statistical analysis, are required to obtain representative I-V or current density-voltage (J-V) curves (**Fig. 15g**).[232, 233]

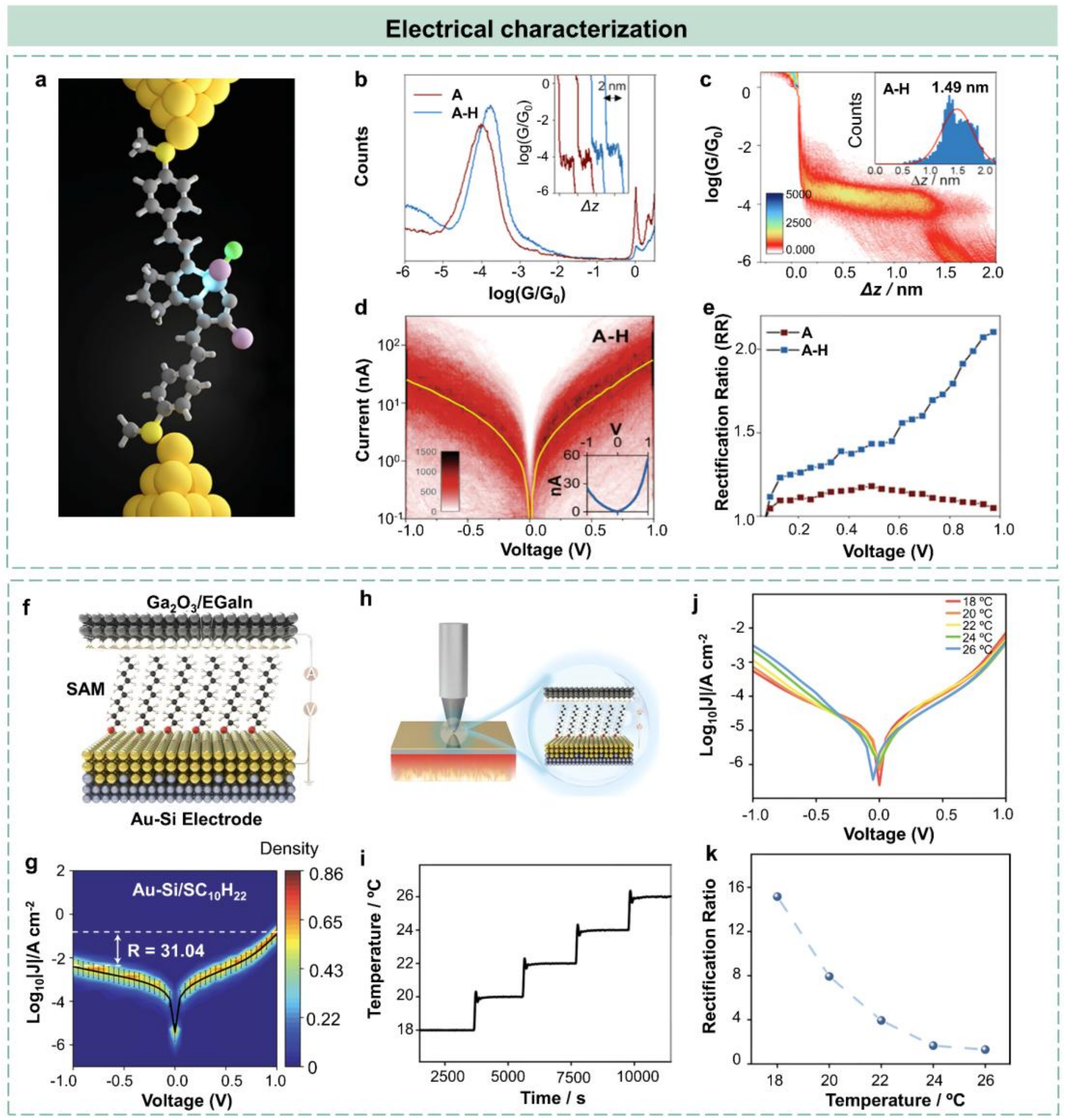

**Figure 15. Electrical characterization. (a)** Molecular structures and device architecture. **(b)** 1D conductance histograms and typical traces for A (brown) and A-H (blue). **(c)** 2D conductance histograms of A-H with corresponding distributions of stretching distances shown in the insets. (**d**) 2D histograms of the *I-V* characteristics for A-H. **(e)** Variation of the RR as a function of the applied voltage. Reproduced with permission Ref. [187]. Copyright 2023, American Chemical Society. **(f)** Schematic of the ideal Au–Si/SAM//$Ga_2O_3$/EGaIn architecture. **(g)** Heat maps of the semilog$_{10}$|*J*|(V) curves for the molecular junction. **(h)** Schematic diagram of the experimental setup for temperature-dependent experiment. **(i)** Real-time temperature curves corresponding to the control of the junction temperature. **(j)** Log$_{10}$|*J*|(V) curves of junctions measured at 18, 20, 22, 24, and 26 °C. **(k)** RR plotted as a function of temperature. Reproduced with permission Ref. [117]. Copyright 2024, American Chemical Society.

Recently, transient current measurements and pulsed-bias techniques have attracted increasing attention.[49] By applying rapidly varying voltage pulses, molecular responses on picosecond or faster timescales can be captured, revealing non-equilibrium charge transport dynamics. Compared with steady-state I-V characterization, these methods better approximate real operational conditions, particularly under high-frequency or non-equilibrium scenarios. When combined with time-resolved spectroscopy or in situ microscopy, they hold promise for uncovering how dynamic conformational changes and interfacial states modulate rectification mechanisms.[234] Temperature-dependent conductance measurements are another essential tool for elucidating transport mechanisms (**Fig. 15h**).[117, 235] By employing real-time temperature control to ensure precise thermal stability (**Fig. 15i**), I-V curves can be recorded across a specific range (**Fig. 15j**). This approach allows differentiation between thermally activated tunneling, coherent tunneling, and multi-step hopping, providing crucial parameters for theoretical modeling. More importantly, evaluating the RR as a function of temperature further validates the thermal stability of the device performance (**Fig. 15k**). Notably, device stability and reliability are equally important.[236, 237] Repeated I-V cycling and prolonged bias tests assess junction durability and current drift. For instance, EGaIn-based molecular junctions can sustain hundreds to thousands of repeated scans with stable output, making them a widely adopted platform for molecular rectification studies.[238] Single-molecule break junctions, though less stable, offer unique advantages in statistical reproducibility, enabling intrinsic transport properties to be discerned.

In summary, dynamic break junctions and static device measurements each present distinct benefits. The former offers high-throughput, tunable structures for statistically significant data, while the latter provides stable, integrated architectures for practical performance evaluation. Together, they establish a comprehensive framework for probing nonlinear charge transport in molecular systems and provide a robust foundation for the rational design and optimization of functional single-molecule rectifiers.

### 5.3 Electronic Structure Characterization

The rectification behavior of molecular junctions fundamentally arises from the relative alignment between molecular orbital energies and the Fermi level of the electrodes. Accordingly, electronic structure characterization plays a central role in the study of molecular rectifiers. Precise determination of the HOMO-LUMO gap, orbital ordering, and their evolution under device operating conditions provides direct insights into charge transport mechanisms and informs molecular design, interface engineering, and device optimization.

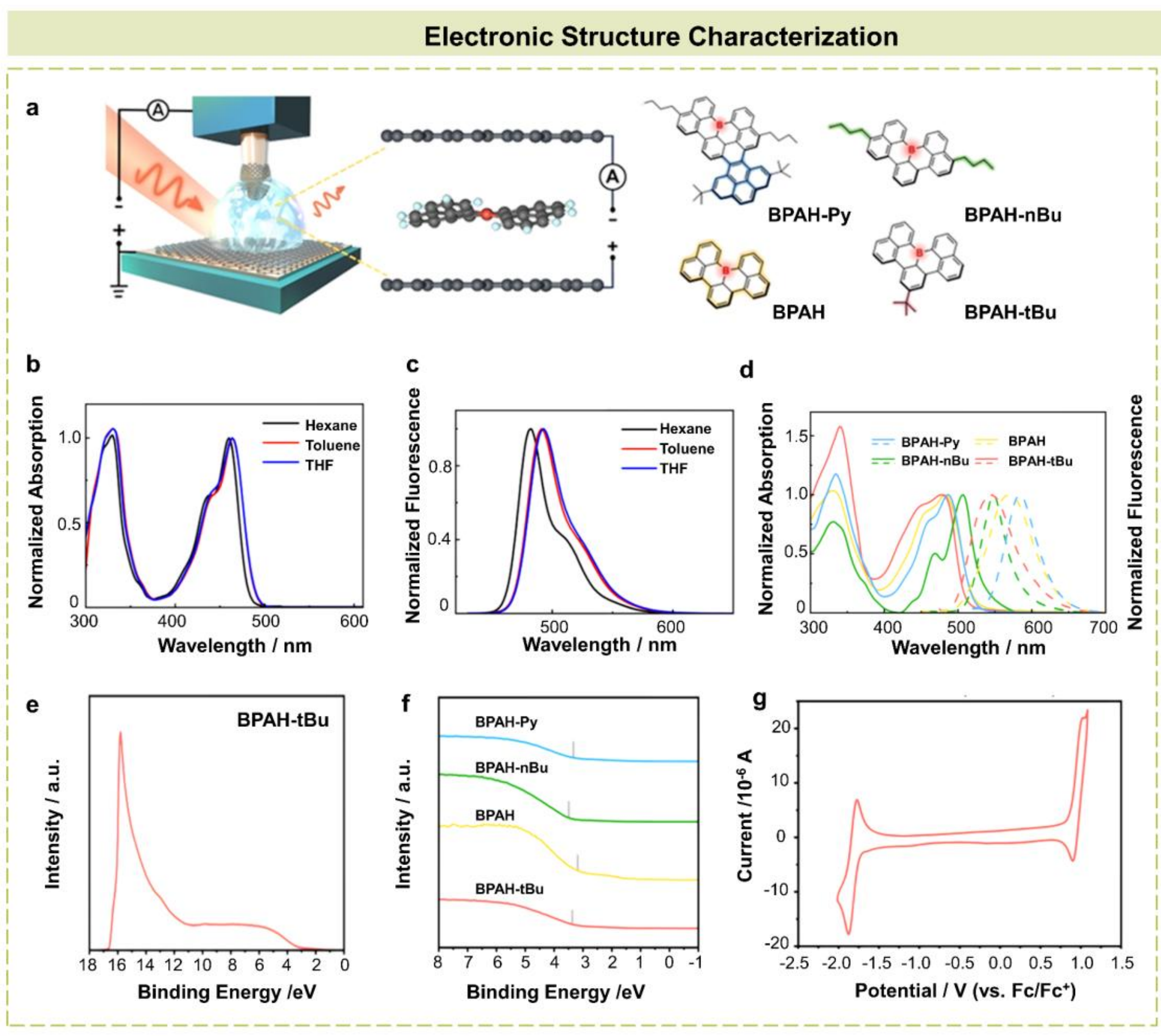

**Figure 16. Electronic structure characterization. (a)** Schematic diagram of the boron-doped single-molecule van der Waals (BPAHs M-2D-vdWHs) diode alongside the chemical structures utilized as functional building blocks. **(b)** Ultraviolet-visible absorption spectroscopy (UV-Vis) absorption and **(c)** fluorescence spectra of BPAH-tBu in solution. **(d)** UV/vis absorption (solid) and fluorescence spectra (dot) of BPAHs in pure film. **(e)** Ultraviolet photoelectron spectroscopy (UPS) of BPAH-tBu M-2D-vdWHs after the XPBJ operation. **(f)** Valence band spectra of BPAHs M-2D-vdWHs. The gray lines indicate the onset energy of the HOMOs. (**g**) Cyclic voltammograms (CV) of BPAH-tBu M-2D-vdWHs. Reproduced with permission Ref. [56]. Copyright 2025, Wiley-VCH GmbH.

Conventional methods, including ultraviolet-visible absorption spectroscopy (UV-Vis), ultraviolet photoelectron spectroscopy (UPS), and cyclic voltammetry (CV), constitute the foundational and complementary toolkit for probing molecular electronic structures.[239-241] A comprehensive characterization of a boron-doped single-molecule van der Waals diode (**Fig. 16a**) exemplifies the synergy of these techniques. UV-Vis spectroscopy measures the absorption of photons in the UV and visible ranges, providing optical gap information. As shown in **Fig. 16b-d**, UV-Vis absorption and fluorescence spectra in both solution and pure film states reveal the impact of molecular packing on electronic transitions. Peak positions reflect the energy difference between HOMO and LUMO levels, while peak intensities correlate with chemical bond types and functional groups, offering additional insight into conjugation and molecular architecture. However, UV-Vis probes the ensemble average in solution or film, often failing to capture the microenvironment of solid-state molecular junctions. UPS, in contrast, directly measures the HOMO level of molecules and the work function of electrodes, enabling evaluation of interfacial energy level alignment and the presence of Fermi-level pinning. For instance, the UPS and valence band spectra of BPAHs (**Fig. 16e-f**) enable the precise determination of onset energies and interfacial energy level alignment. By combining UPS-determined HOMO energies with the optical gaps from UV-Vis, the LUMO level can be indirectly estimated, providing a practical reference for understanding molecule-electrode energy matching. Besides, CV offers a convenient and versatile means to infer molecular energy levels via redox potentials (**Fig. 16g**). Yet, these measurements are sensitive to the reference electrode and solution environment, and thus should be interpreted primarily as solution-phase energy estimates rather than direct analogues of solid-state junction alignment. Collectively, these conventional methods deliver a broad view of molecular electronic structure but remain limited in resolving single-molecule heterogeneity or interfacial subtleties.

To address these limitations, transport-based characterization techniques have been increasingly adopted in molecular rectifier studies. Transition voltage spectroscopy (TVS) analyzes the *I-V* characteristics to extract the bias voltage associated with the dominant molecular orbital, which serves as an approximate measure of the tunneling barrier height.[242, 243] As illustrated in the study of Pyr-Cn SAM junctions (**Fig. 17a)**, the transition from Simmons tunneling (rectangular barrier) to Fowler-Nordheim tunneling (triangular barrier) can be precisely identified via the minima in F-N plots (**Fig. 17b**).[244] Specifically, the lower $V_T$ values observed for Pyr-C4S2 compared to Pyr-C12 (**Fig. 17c**) reveal that the tunneling barrier is highly sensitive to the molecular orientation and packing density of the pyrenyl moieties rather than just the chemical nature of the alkyl chains. Inelastic electron tunneling spectroscopy (IETS) further reveals vibrational contributions during electron tunneling, enabling identification of transport mechanisms such as resonant tunneling, off-resonant tunneling, or multi-step hopping.[180] For a diblock molecular diode (**Fig. 17d**), while the inherent conductance asymmetry can obscure standard features in the *G*-*V* curve, clear antisymmetric peaks emerge in the IET spectra (**Fig. 17e**). Crucially, the observation that phonon modes occur at identical energies for both bias polarities, despite significant differences in current magnitude demonstrates that the potential profile does not shift the vibrational excitation levels, confirming that tunneling dominates the transport (**Fig. 17f**). Kelvin probe force microscopy (KPFM) offers nanoscale mapping of local contact potential differences, providing indirect information on interfacial dipoles, local charge rearrangements, and molecular orientation, all of which modulate rectification behavior.[115] Combining conductive-probe AFM (CP-AFM) with KPFM (**Fig. 17g**) enables the spatial mapping of surface potentials (**Figs. 17h** and **i**) and demonstrates how the work function of the bottom electrode can be tuned by the molecular length (**Fig. 17j**). Such quantitative data on interfacial energy level alignment is critical for understanding the electrostatic environment that governs rectification. Compared with conventional optical or electrochemical approaches, TVS, IETS, and KPFM deliver transport-relevant quantitative or spatially resolved data, particularly valuable for investigating non-equilibrium states or local heterogeneity.

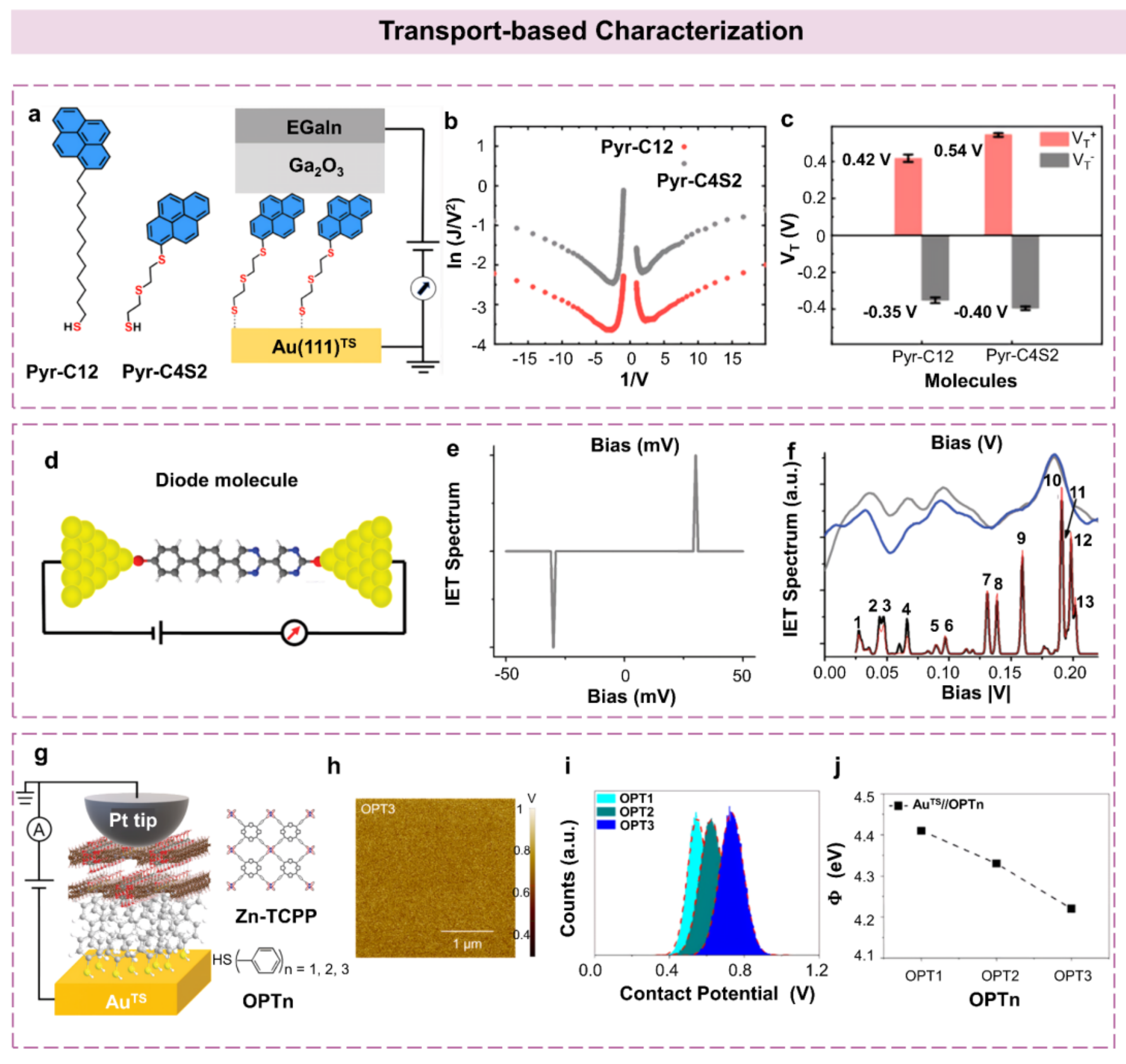

**Figure 17. Transport-based characterization. (a)** Schematic illustration of two-terminal SAM junction and the chemical structures of Pyr-C12 and Pyr-C4S2. **(b)** Fowler-Nordheim (F-N) plots for Pyr-C12 and Pyr-C4S2 SAMs. **(c)** Derived transition voltages at positive $V_T^+$ and negative $V_T^-$ bias for the corresponding molecular junctions. Reproduced with permission Ref. [244]. Copyright 2023, American Chemical Society. **(d)** Depiction of the diblock molecule designed to behave as a molecular diode via its asymmetric electronic structure. **(e)** Antisymmetric peaks in the inelastic electron tunneling spectrum (IETS) indicating molecular vibrational signatures. **(f)** Subtracted IET spectra obtained by taking the derivative of the IET signal and comparing the experimental curves at positive (blue) and negative (gray) bias with calculated spectra (black and red). Reproduced with permission Ref. [180]. Copyright 2011, American Chemical Society. **(g)** Schematic of the electronic measurement setup for molecular heterojunctions using CP-AFM with 2D Zn-TCPP nanosheets on OPTn SAMs. **(h)** KPFM images of OPT3 SAMs on $Au^{TS}$ substrates. **(i)** Histograms of the contact potential for OPTn (n = 1, 2, 3) SAMs. **(j)** Correlation between the work function (Φ) and the contact potential across the OPTn series with varying molecular lengths. Reproduced with permission [115]. Copyright 2025, Wiley-VCH GmbH.

Nevertheless, current electronic structure characterization approaches remain constrained. UV-Vis, UPS, and CV are typically performed under idealized solution or vacuum conditions, while IETS and KPFM measurements are conducted under environments that differ from actual device operation, limiting their ability to capture single-molecule variability and realistic interfacial behavior. Moreover, data interpretation in IETS and KPFM is heavily reliant on theoretical modeling. Future advancements should emphasize in situ, multimodal approaches that couple electronic structure characterization with simultaneous electrical transport measurements, enabling direct observation of molecular junctions under working conditions.[54, 245] Such strategies promise to bridge the gap between idealized experiments and practical device operation, offering a more accurate and comprehensive understanding of the electronic structure and transport mechanisms governing molecular rectification.

# 6 Theoretical Modelling and Simulation

First-principles simulations represent a class of approaches that are directly rooted in the fundamental laws of quantum mechanics and statistical mechanics, without relying on empirical parameters. In the context of molecular rectifiers, these methods not only allow the calculation of electronic structures and rectification performance, but also unravel the microscopic mechanisms underlying I-V responses. The most established theoretical framework builds on the combination of DFT and NEGF methods.[39, 246] This section highlights the key computational steps and applications of this framework in molecular rectifier research. Accurate electronic structure inputs are first established using DFT, followed by NEGF-based simulations of quantum transport under nonequilibrium conditions, thereby providing a coherent theoretical bridge between molecular-level and device-level descriptions.

Accurate electronic structure input is a prerequisite for molecular transport simulations. DFT, as the mainstream method for molecular-scale electronic structure calculations, uses electron density rather than many-electron wavefunctions as its fundamental variable.[246-249] This ensures a reasonable balance between accuracy and computational efficiency, substantially reducing the cost of calculations. Within the Born-Oppenheimer approximation, the complex many-electron problem is mapped onto a system of non-interacting electrons moving in an effective potential, and the Kohn-Sham equations are solved to obtain the ground-state electron density and total energy of the system.[250] Owing to its balance of accuracy and efficiency, DFT has become an indispensable theoretical foundation in the study of molecular rectifiers.

In molecular rectifier studies, the most common application of DFT is the calculation of intrinsic molecular electronic structures.[101, 142] By determining the energy levels and spatial distributions of the HOMO and the LUMO, DFT can uncover the electronic origin of rectification effects (**Fig. 18**a).[251, 252] This is particularly critical for understanding donor-acceptor (D-A) type rectifying systems, whose rectification behavior arises directly from the energy splitting and asymmetric spatial distribution of frontier orbitals, which can be qualitatively predicted by DFT calculations. In addition, key physical quantities such as dipole moments and polarizabilities can be computed, providing insight into molecular responses under external electric fields, which are intimately linked to asymmetric rectification.[96] Moreover, charge density difference analysis and Bader charge decomposition enable quantitative evaluation of molecule-electrode charge transfer and interface dipole formation, clarifying the role of MIGS in level alignment and transport performance.[99, 253, 254]

The choice of exchange-correlation functional is central to reliable predictions. The generalized gradient approximation (GGA, e.g., PBE) remains widely used due to its robustness in structural and band calculations, but systematically underestimates gaps and level alignments.[255, 256] Hybrid functionals (e.g., B3LYP, HSE06), by incorporating a fraction of Hartree-Fock exchange, improve band-gap predictions, while long-range corrected functionals (e.g., ωB97XD, CAM-B3LYP) perform better in describing charge transfer and polarization effects.[257-261] For systems involving transition metals or strong electron correlation, the DFT+U approach is necessary to capture localized *d*/*f* orbital behavior, whereas time-dependent DFT (TD-DFT) provides a natural framework for excited states and dynamical processes.[262, 263] Commonly used computational platforms include Gaussian 16 (molecular levels and spectra), VASP (Vienna Ab initio Simulation Package, interface modeling), Quantum ATK (integrated electronic structure and transport), and SIESTA (efficient for large-scale systems).[264-267]

Despite its central role, DFT faces notable limitations. Kohn-Sham eigenstates are mathematical constructs rather than true physical observables, leading to systematic deviations of up to ~1 eV in predicting ionization potentials (IP) and electron affinities (EA) compared with experiments or higher-level theories such as GW.[268] This uncertainty strongly impacts the alignment between molecular energy levels and electrode Fermi energies, which in turn influences tunneling resonances, current magnitude, and RRs. More accurate methods such as GW or coupled-cluster theory (CCSD) offer superior predictions but remain computationally prohibitive for realistic device scales.[269, 270] Emerging strategies, including machine learning-assisted functional optimization, high-throughput screening, and GPU acceleration, are beginning to alleviate the accuracy-efficiency trade-off.[271] Meanwhile, multiscale QM/MM coupling schemes provide a pathway toward realistic large-scale simulations.[272] These advances are driving the transition from atomic-scale electronic descriptions to full device-level modeling, laying the theoretical groundwork for rational design and optimization of molecular rectifiers.

In contrast to DFT, which primarily focuses on molecular and interface electronic structures, the DFT-NEGF framework establishes a comprehensive tool for studying nonequilibrium transport in molecular rectifiers under bias.[273] By integrating the electronic structure from DFT with the NEGF, researchers can compute I-V characteristics from first principles and evaluate RRs. Crucially, this approach enables direct correlation between microscopic transport mechanisms and experimental measurements, thus forming a self-consistent loop between theory and experiment.

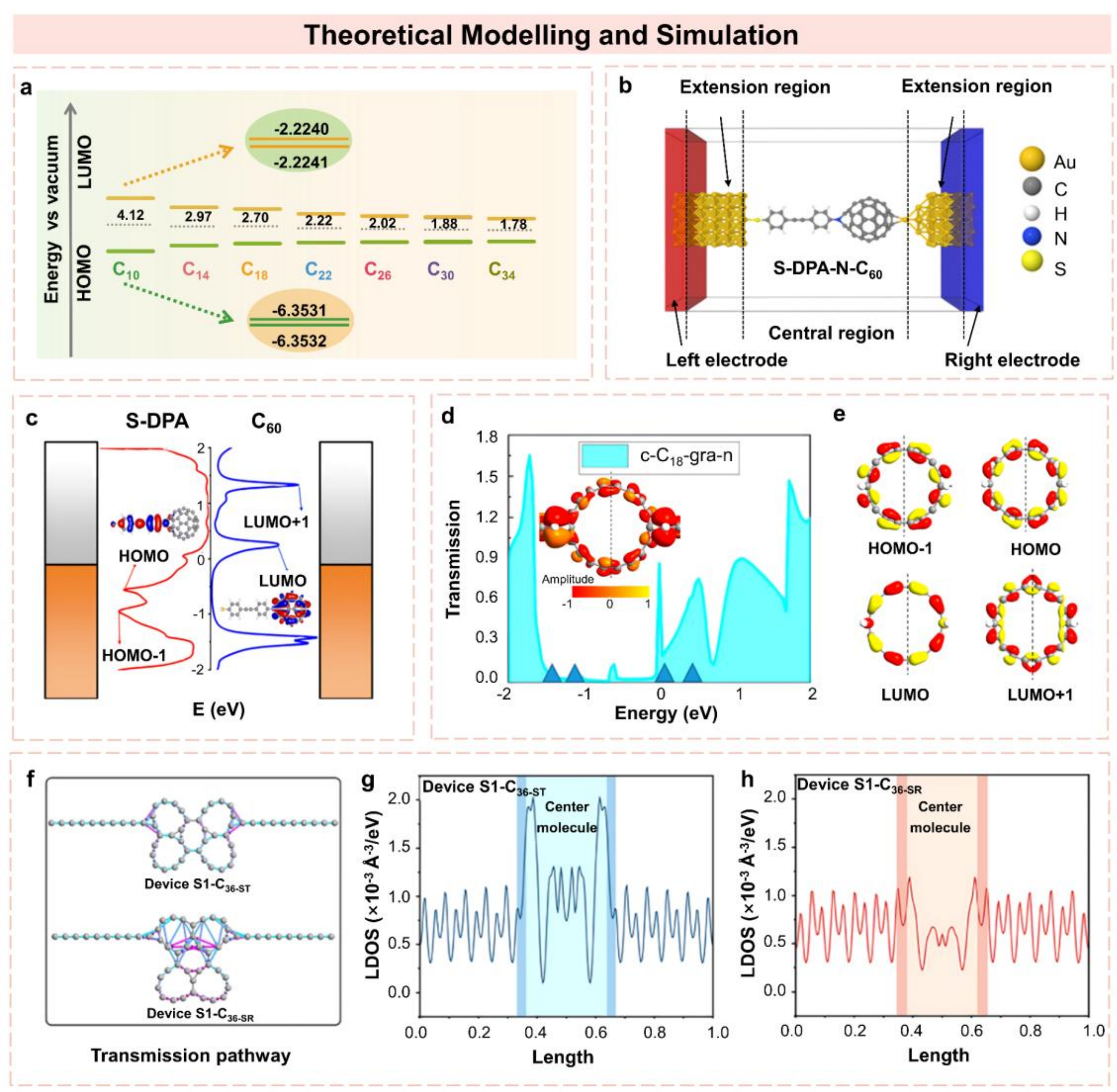

**Figure 18. Density Functional Theory (DFT) and Nonequilibrium Green's Function (NEGF) calculations. (a)** Frontier molecular orbitals energies of Cyclo[n]carbon. The Fermi level is indicated by the gray dashed line. Reproduced with permission Ref. [251]. Copyright 2025, Royal Society of Chemistry. **(b)** Schematic representation of the S-diphenylacetylene (DPA)-N-$C_{60}$ device model, where the red and blue boxes denote the electrode sections. **(c)** Real-space local density of states (RLDOS) for the S-DPA (left red line) and $C_{60}$ (right blue line) moieties at 0 V, with the inset illustrating the HOMO and LUMO orbital isosurfaces. Reproduced with permission Ref. [129]. Copyright 2025, Royal Society of Chemistry. (**d**) Equilibrium transmission spectrum of the c-$C_{18}$-gra-n device. Four dominant molecular projected self-consistent Hamiltonian eigenvalues near the Fermi level are marked by blue triangles, and the Fermi level is uniformly set to 0 eV. The inset displays the transmission eigenstate corresponding to the main transmission peak of the device under equilibrium conditions. **(e)** MPSH states of four molecular frontier orbitals. Reproduced with permission Ref. [275]. Copyright 2020, American Chemical Society. **(f)** Transmission pathways of device S1-$C_{36\text{-ST}}$ (upper) and device S1-$C_{36\text{-SR}}$ (lower) under equilibrium conditions. **(g-h)** LDOS for device S1-$C_{36\text{-ST}}$ **(g)** and device S1-$C_{36\text{-SR}}$ **(h)**. Reproduced with permission Ref. [276]. Copyright 2025, AIP Publishing.

In device modeling, the system is typically partitioned into a left electrode, a central scattering region, and a right electrode.[274] As illustrated by the S-diphenylacetylene (DPA)-N-$C_{60}$ device model (**Fig. 18b**), the central region includes not only the molecule of interest but also several electrode atoms, forming an "extended molecule."[129] This treatment accounts for hybridization between molecular orbitals and electrode states, a key factor in ensuring physical reliability.[133] The computational procedure typically begins with self-consistent DFT calculations to obtain the Hamiltonian and electronic states of the central region. Within the NEGF framework, the electrode self-energy matrices are then introduced to simulate semi-infinite electrodes. This approach allows for the calculation of I-V characteristics and the RR. In specific transport analysis, the real-space local density of states (RLDOS) at zero bias and the corresponding HOMO/LUMO isosurfaces can intuitively reveal the equilibrium electronic distribution characteristics.[129] The spatial asymmetry of the orbital distribution can serve as a preliminary indicator for judging the existence of potential rectification behavior in the system (**Fig. 18**c). To further reveal the microscopic transport mechanism, the evolution of the transmission spectrum T(E,V) with energy can be calculated to determine the relative position between molecular orbitals and the Fermi level of the electrodes. The profile of the transmission spectrum reflects the resonant coupling between molecules and electrodes (**Fig. 18**d).[275] It is worth noting that the molecular orbit dominating charge transport is not always the intrinsic orbital closest to the Fermi level. Combined analysis of the transmission eigenstates of the device and the molecular projected self-consistent Hamiltonian (MPSH) can accurately identify the spatial transport channel that contributes the most to the current. As shown in **Figs. 18**d-e, although the system contains multiple molecular orbitals, orbital projection analysis reveals that the LUMO+1 orbital becomes the core orbital dominating charge transport due to its higher spatial overlap with the electrode states. To further clearly present the internal transport image of electrons in the device, transmission pathway analysis can be introduced to intuitively show the distribution and evolution of current along the molecular skeleton and at the interfaces at the real-space scale **(Fig. 18**f).[276] Meanwhile, the coupling strength between molecules and electrodes can be quantitatively evaluated by analyzing the distribution of the local density of states (LDOS) at the interface **(Figs. 18**g-h). Higher LDOS at the interface usually corresponds to stronger orbital hybridization and better charge injection efficiency, thus significantly improving the overall transport performance of the device. The above multi-dimensional simulation analysis provides a solid theoretical foundation for optimizing the device rectification ratio from the perspectives of molecular structure regulation and interface engineering. More importantly, the orbital evolution and transport mechanism under external bias can be directly visualized, especially the energy shift and spatial redistribution of the orbitals. These results clearly clarify the physical origin and evolution of the molecular rectification effect at the atomic scale.

Furthermore, this methodology enables a systematic assessment of how molecular conformation, anchoring groups, electrode materials, and contact geometries impact rectification performance.[95, 176, 253] Crucially, the DFT-NEGF framework can capture QI effects and provide a robust platform to explore external-field-induced orbital evolutions, such as those triggered by gate modulation or mechanical strain.[136, 137] Several computational platforms have implemented the DFT combined with NEGF approach, including TranSIESTA, QuantumATK, OpenMX, and GPAW, all of which support self-consistent transport calculations under bias and facilitate detailed modeling of molecule-electrode coupling.[266, 277-279]

Taken together, DFT-NEGF has become a central theoretical bridge connecting molecular electronic structures to nonequilibrium transport phenomena, offering powerful insights into device design and interface engineering. Yet challenges remain, including high computational cost, convergence difficulties, and insufficient treatment of strong correlation and inelastic scattering. Future directions will focus on improving computational efficiency for large and multi-physics systems, integrating advanced many-body Green's function methods, developing models for strongly correlated systems, and leveraging machine learning for transport prediction.[280] These developments will advance the field from mechanistic understanding at the single-molecule level toward the predictive design of high-performance molecular rectifiers.

# 7 Comparison of Representative Molecular Rectifiers

In recent years, molecular rectifiers have witnessed remarkable progress in performance enhancement, innovative regulation strategies, and advances in fabrication and characterization techniques. As shown in **Table 1**, this section systematically reviews a series of representative systems with excellent performance and diverse mechanisms, and provides a multidimensional comparison of them. This not only deepens the understanding of the working mechanisms of molecular rectifiers but also provides valuable experience for the optimization and integration of future devices.

From the core performance indicator RRs of molecular rectifiers, the current maximum RR can reach the order of $10^5$-$10^8$ (such as $HS(C_{12})_{11}BPY–C_n–C–BPY$ and $C_6$-DNTT systems), approaching or even surpassing some traditional inorganic diodes, highlighting the great potential of molecular rectifiers. These high-performance devices are predominantly based on SAM platforms and grounded in the KBW model, where synergistic structural and interfacial engineering plays a decisive role. In contrast, single-molecule devices generally exhibit lower RRs. Nevertheless, they offer unique advantages for theoretical exploration and functional design, particularly for investigating external-field modulation such as temperature, gate bias, and quantum interference.

From the perspective of design strategies, the AR and KBW models remain the theoretical cornerstones of molecular rectification. More than half of the reported systems rely on the combine effect of molecular structural and interfacial engineering, demonstrating their versatility and effectiveness. Meanwhile, quantum interference and multi-physics-field modulation, through electric fields, optical, temperature, or chemical environments are rapidly emerging as powerful means to transcend traditional rectification mechanisms. For instance, field-induced conformational rearrangements can activate quantum interference under opposite bias, generating direction-dependent transport. Photo-switchable molecules can achieve reversible on-off rectification under UV/visible irradiation. Thermal and environmental modulation further reveal the influence of thermally activated transport and interfacial ionic reorganization. Together, these developments highlight multidimensional synergistic regulation as a key pathway toward programmable molecular rectifiers.

In terms of electrode selection, noble metals such as Au, Ag, and Pt still dominate, yet the application of alternative materials is steadily increasing. Au/EGaIn electrodes account for nearly half of recent SAM-based devices. The soft and conformal contact of EGaIn effectively prevents disruption of molecular layers during metal deposition, making it the preferred choice for SAM-based rectifiers. Carbon-based electrodes, including graphene and carbon nanotubes, have attracted attention due to their ability to suppress MIGS, preserve molecular orbital integrity, and enhance interfacial stability. In addition, semiconductor-metal hybrid electrodes such as N-type Si/Au further expand experimental approaches and offer opportunities for integration with conventional semiconductor technology.

On the fabrication and characterization front, dynamic break-junction methods such as MCBJ and STM-BJ complement static SAM techniques. While SAM dominates in devices with high RR, break-junction approaches remain indispensable for single-molecule studies. Characterization increasingly involves multidimensional strategies, combining structural probes (NMR, XRD, AFM), electronic structure analysis (XPS, UPS, NEXAFS), and electrical measurements (I-V, J-V-T, TVS) to ensure reliable performance evaluation and mechanistic understanding. In parallel, theoretical simulations based on DFT, DFT combined with NEGF, and molecular dynamics continue to provide critical insights into level alignment, charge transport, and external-field effects, effectively bridging experimental and theoretical efforts.

Overall, molecular rectifier research in recent years has demonstrated a close interplay between theory and experiment, yielding significant advances in performance optimization, multi-mechanism regulation, and fabrication-characterization methodologies. However, challenges still exist, such as the difficulty for individual molecules to achieve high RR, insufficient device stability, and the difficulty of large-scale fabrication of carbon-based electrodes. In the future, if breakthroughs can be made in device stability, scalable fabrication, and integration with semiconductor platforms, molecular rectifiers are expected to truly move toward practical applications.

**Table 1. Comparison of Representative Molecular Rectifiers**

| Molecule | RR (Max) | Electrodes | Basic Model | Modulation Strategies | Fabrication Techniques | Characterization Techniques | Computational Methods | Reference |
|---|---|---|---|---|---|---|---|---|
| **$C_6$-DNTT** | $5.0 \times 10^8$ | Pt/Ti | — | Molecular structure<br>Interface | Monolayer | AFM, ARXPS, CA, J-V, J-V-T, KPFM, TEM, TVS, UPS, UV-Vis, XRD | — | [26] |
| **T-exTTF** | $5.0 \times 10^4$ | Au/EGaIn | — | Molecular structure<br>Interface | SAM | CV, Ellipsometry, HRMS, J-V, NMR, UPS, XPS, XRD | DFT (Gaussian)<br>DFT-NEGF (Quantum ATK) | [197] |
| **phenyl benzoate** | $6.4 \times 10^3$ | Graphene/Graphene | — | Quantum interference<br>Chemical environment<br>Temperature | Carbon-based | IETS, I-V, I-t | DFT (Gaussian)<br>DFT-NEGF (Quantum ATK) | [48] |
| **Py/NDI** | 16 | Au/Au | AR | Quantum interference<br>Molecular structure | MCBJ<br>STM/BJ | Flicker noise, HRMS, I-V, MS, NMR, SEM, UV-Vis, XRD | DFT(Gaussian)<br>DFT-NEGF (Quantum ATK) | [49] |
| **$HSC_{15}Fc–C≡C–Fc$** | $1.0 \times 10^5$ | Pt/EGaIn | KBW | Molecular structure<br>Interface | SAM | CV, DSC, HRMS, J-V, NEXAFS, NMR, UPS, XPS | MD (Gromacs) | [281] |
| **$S(CH_2)_{10}OPhBr_5$** | 40 | Ag/EGaIn | KBW | Molecular structure<br>Interface | SAM | AFM, EIS, J-V, J-V-T, NEXAFS, NMR, UPS, UV-Vis, XPS | DFT(Gaussian) | [114] |

| **BPAH-tBu** | 457 | Graphene/Graphene | — | Molecular structure<br>Interface | XPBJ | CV, HRMS, I-V, NMR, RS, UPS, UV-Vis, XRD | DFT (Gaussian)<br>DFT-NEGF (Quantum ATK)<br>TD-DFT | 56 |
|---|---|---|---|---|---|---|---|---|
| **IT-6** | 3.63 | Au/Au-Tip | — | Molecular structure<br>Chemical Environment | STM-BJ | CV, HRMS, I-V, NMR, UV-Vis | DFT (Gaussian)<br>TD-DFT | 27 |
| **Fc-BPT** | $1 \times 10^3$ | Au/EGaIn | AR | Molecular structure<br>Interface | SAM | CV, J-V, NEXAFS, XPS | DFT (Gaussian) | 235 |
| **HBQ** | $1 \times 10^4$ | Au/EGaIn | AR | Molecular structure<br>Interface<br>Temperature | SAM | AFM, CV, HRMS, J-V, J-V-T, NEXAFS, NMR, STM, UPS, UV-Vis | DFT (Gaussian) | 282 |
| **$S(CH_2)_{11}S$–BTTF** | 124 | Au/EGaIn | KBW | Molecular structure<br>Interface<br>Temperature | SAM | ARXPS, CV, J-V-T, NEXAFS, UPS, XPS | — | 151 |
| **$S(CH_2)_2CONR^1R^2$** | 11 | Au/EGaIn | KBW | Molecular structure<br>Interface | SAM | ARXPS, CA, HRMS, I-V, J-V, NMR, XPS | MD (LAMMPS)<br>DFT (Gaussian) | 236 |
| **Ru-DAE** | 61 | Graphene/Graphene | — | Electric Field<br>Optical Field | Carbon-based | HRMS, I-V, NMR, SEM, UV-Vis | DFT (Gaussian, VASP)<br>DFT-NEGF (Nanodcal) | 28 |
| **$S(CH_2)_{11}$BIPY–$CoCl_2$** | 82 | Au/EGaIn | KBW | Molecular structure<br>Interface | SAM | CV, J-V-T, J-V, UPS, XPS | DFT (Gaussian) | 116 |

| **Cu–BTEC** | $3.1 \times 10^5$ | Au/EGaIn | KBW | Molecular structure<br>Interface<br>Temperature | SAM | CA, CV, EPR, FE-SEM, I-V, I-V-T, IRRAS, UV-Vis, XPS, XRD | — | 283 |
|---|---|---|---|---|---|---|---|---|
| **Fc-PTM** | 99 | Au/EGaIn | AR | Molecular structure<br>Interface | SAM | AFM, CV, FT-IR, J-V, NEXAFS, NMR, SEM, UV-Vis, UPS, XPS | DFT(Gaussian, SIESTA) | 238 |
| **$SC_{11}PAH$** | 172 | Ag/EGaIn | KBW | Molecular structure<br>Interface | SAM | AFM, CA, HRMS, J-V, J-V-T, NMR, UPS, UV-Vis, XPS | — | 142 |
| **$S(CH_2)_{11}CO_2H$** | 100 | Au/EGaIn | KBW | Molecular structure<br>Interface<br>Optical Field<br>Chemical Environment | SAM | AFM, J-V, KPFM, XPS | DFT (Gaussian) | 160 |
| **$HS(CH_2)_{15}Fc–C{\equiv}C–Fc$** | $6.3 \times 10^5$ | Pt/EGaIn | KBW | Molecular structure<br>Interface | SAM | AFM, CV, DSC, HRMS, J-V-T, J-V, NEXAFS, NMR, UPS, XPS | MD (NAMD) | 24 |
| **1,8-nonodiyne** | $4 \times 10^3$ | N-type Si/Au | | Molecular structure<br>Interface | STM-BJ | AFM, I-V, STM, XPS | DFT (Gaussian) | 183 |
| **DPE-2F** | 620 | Au/Au | AR | Molecular structure<br>Electric Field | MCBJ | I-V, SEM | DFT (ADF)<br>DFT-NEGF | 143 |
| **pyridinoparacyclophane** | 2.63 | Au/Au-Tip | AR | Molecular structure<br>Interface | STM-BJ | I-V, NMR, PL, TVS, UV-Vis | DFT (Gaussian)<br>DFT-NEGF<br>(Quantum ATK) | 96 |

| **hetero *π*-stacked complex 1·(2·3):2** | 10 | Au-Tip | AR | Molecular structure<br>Interface | STM-BJ | I-V, NMR, STM, XRD | DFT (Gaussian)<br>DFT-NEGF | 102 |
|---|---|---|---|---|---|---|---|---|
| **$S(CH_2)_{11}MV^{2+}I_2^{-}$** | $2.5 \times 10^4$ | Ag/EGaIn | KBW | Molecular structure<br>Interface | SAM | ARXPS, CV, HRMS, J-V, J-V-T, NEXAFS, NMR, UPS, UV-Vis, XPS | DFT (Gaussian) | 191 |
| **$HSC_{11}$BIPY–C≡C–BIPY** | $8.26 \times 10^6$ | Pt/Pt | KBW | Molecular structure<br>Interface | — | — | DFT (Quantum ATK)<br>NEGF-DFT (Quantum ATK) | 108 |
| **pyrene-$(CH)_3$-(CH2)$_3$-$(CH)_3$-benzene** | 93.8 | Au/Au | AR | Molecular structure | — | — | DFT (VASP)<br>NEGF-DFT (Quantum ATK) | 98 |
| **2,2'-bipyridyl** | 388.6 | Ag/Ag | KBW | Molecular structure<br>Interface | — | — | DFT (Quantum ATK)<br>NEGF-DFT (Quantum ATK) | 93 |
| **pyrimidinyl-triphenyl** | 1.65 | Au/Au | AR | Molecular structure | — | — | DFT (SIESTA)<br>NEGF-DFT (TranSIESTA） | 91 |
| **2,4-TA** | 120.2 | Au/Au | | Quantum interference<br>Electric Field | — | — | DFT (Gaussian)<br>DFT-NEGF (Quantum ATK) | 284 |
| **N-Phenylbenzamide** | 16.67 | Au/Au | AR | Molecular structure | — | — | DFT(Gaussian)<br>NEGF-DFT (TranSIESTA) | 90 |

# 8 Summary and Outlook

As a promising avenue to surpass the physical limits of conventional silicon-based technologies in the post-CMOS era, molecular rectifiers consistently demonstrate unique advantages in ultimate device scaling and potential applications. Despite continuous performance improvements, whether they can replace silicon components at the standard device level remains highly uncertain. This uncertainty primarily stems from the formidable challenges in realizing molecular devices that combine high RRs, robust stability, and integrability. Fundamentally, these challenges can be distilled into three core scientific questions: How can highly controllable and structurally stable devices be constructed experimentally? How can the intrinsic relationship between molecular structure and electronic transport behavior be accurately elucidated theoretically? And how can molecular rectifiers be effectively integrated into complex electronic systems at the application level? This section provides a critical assessment of current bottlenecks across these three dimensions, along with strategies and future perspectives aimed at driving sustained progress in this field.

**1) Experimental efforts: constructing robust and controllable molecular rectifiers**

The precise fabrication of nanogap electrodes and the controlled assembly of molecule-electrode interfaces remain key bottlenecks limiting device performance. As device dimensions approach the single-molecule scale, rectification characteristics become extremely sensitive to structural fluctuations and interfacial perturbations, placing stringent demands on electronic coupling and chemical stability at the interface. Although break junctions, scanning probe methods, and self-assembled monolayers have enabled initial device construction, issues such as poor configurational uniformity and limited reproducibility of electrical responses remain pervasive. Moreover, bias-induced molecular rearrangements, electrode atom migration, and dynamic evolution of interfacial states can lead to nonlinear fluctuations and instability during device operation. Addressing these challenges will require nanoscale fabrication techniques with atomic precision, the use of highly selective anchoring groups, and the adoption of low-diffusivity electrode materials to enhance interfacial stability and charge injection efficiency. In addition, the development of in situ multimodal characterization techniques is essential to monitor interfacial dynamics in real time during device operation. Bioinspired materials such as DNA scaffolds and mechanically interlocked molecules, which exhibit both self-repair capabilities and excellent conductivity, may also offer alternative platforms for building next-generation, high-stability molecular devices.

**2) Theoretical insights: understanding and predicting rectification mechanisms**

Theoretical modeling plays a critical role in the design, mechanistic understanding, and performance prediction of molecular rectifiers. While the DFT combined with NEGF formalism has become the standard modeling framework, it still falls short in capturing key physical mechanisms such as energy-level alignment, MIGS, and interfacial charge reorganization. In particular, the systematic underestimation of energy gaps by generalized gradient approximation functionals like PBE can lead to inaccurate predictions of rectification behavior. To enhance predictive accuracy and broaden applicability, it is essential to incorporate higher-level methods such as hybrid functionals, the GW approximation, and TD-DFT. Multiscale modeling frameworks that combine ab initio calculations with molecular dynamics and machine learning approaches are also crucial to balance computational efficiency with the complexity of real systems. Developing general transport models that can account for realistic environments, including solvents, electric fields, and mechanical stress, will further improve the complementarity between theory and experiment and enhance mechanistic interpretability.

**3) Toward integration and commercialization**

Although experiments have demonstrated excellent rectification ratios and tunability in molecular rectifiers, their transition toward practical applications is hindered by three key barriers. First, their operational stability and environmental tolerance remain insufficient for long-term reliable use. Second, current fabrication techniques are not yet compatible with mainstream semiconductor processes, which limits the scalable and controllable construction of device arrays. Third, the typical output currents remain in the picoampere to nanoampere range, which is insufficient to drive conventional electronic circuits. One potential pathway toward commercialization involves embedding molecular rectifiers within heterogeneous platforms that integrate CMOS devices, two-dimensional materials, quantum dots, or flexible electronics. This approach could harness their low power consumption, high integration density, and programmable functionality at the system level. Customizing molecular rectifiers for specialized applications such as flexible sensing, bio-interfaces, and neuromorphic computing could also offer viable entry points for commercial deployment. Establishing standardized testing protocols, reproducible manufacturing processes, and unified performance evaluation frameworks will be critical for bridging the gap between research and industrial implementation.

In summary, the transition of molecular rectifiers from scientific prototypes to engineered applications requires not only close integration between theory and experiment but also coordinated advances across disciplines and scales. Despite the significant challenges ahead, the ultra-small dimensions and low-power operation of molecular rectifiers highlight their profound potential in next-generation nanoelectronic systems. With continued progress in advanced characterization and integrated manufacturing technologies, molecular rectifiers are poised to play an increasingly substantive role in future information systems, potentially catalyzing a paradigm shift in microelectronics.

## Acknowledgments

We would like to acknowledge the support from the National Natural Science Foundation of China (NNSFC) (Grant Nos. 52501308), the Key Research and Development Plan of Shandong Province (Grant Nos. 2025CXGC020107 and 2021SFGC1001), the Key Technology Research and Development Program of Shandong Province (No. 2025CXGX010406) and the Natural Science Foundation of Shandong Province (No. ZR2025QC1106). This work is also supported by the Special Funding in the Project of the Taishan Scholar Construction Engineering and the program of Jinan Science and Technology Bureau (2020GXRC019) as well as new material demonstration platform construction project from Ministry of Industry and Information Technology (2020-370104-34-03-043952-01-11).

## Conflict of Interest

The authors declare no competing financial interest.

# Data Availability Statement

The data that support the findings of this study are available from the corresponding author upon reasonable request.